\documentclass[%
 aip,
 amsmath,amssymb,
 reprint,%
]{revtex4-1}

\usepackage{graphicx}
\usepackage{dcolumn}
\usepackage{bm}
\usepackage{lipsum}

\usepackage[utf8]{inputenc}
\usepackage[T1]{fontenc}
\usepackage{mathptmx}
\usepackage{xcolor}
\usepackage{mathrsfs}

\begin{document}

\preprint{AIP/123-QED}

\title{Flow dynamics of wormlike micellar solutions through a model porous media}
\author{Mohd Bilal Khan}
\author{C. Sasmal}%
 \email{csasmal@iitrpr.ac.in}
\affiliation{ 
Soft Matter Engineering and Microfluidics Lab, Department of Chemical Engineering, Indian Institute of Technology Ropar, Punjab, India-140001.
}



\date{\today}

\begin{abstract}
 The flow of viscoelastic wormlike micellar solutions (WLMs) in porous media is encountered in many practical applications like enhanced oil recovery or groundwater remediation. To understand the flow dynamics of these complex fluids in porous media,  a model porous media consisting of a straight microchannel with micropore throats present in it, is very often used. In this study, we perform an extensive numerical investigation to understand the flow dynamics of wormlike micellar solutions based on the two-species Vasquez-Cook-McKinley (VCM) model for micelles through such model porous media. We find the existence of an elastic instability once the Weissenberg number exceeds a critical value; likewise, it was seen in many prior experimental and numerical studies dealing with polymer solutions. However, for the present case of a WLM solution, we observe that this elastic instability is greatly influenced by the breakage and reformation mechanisms of the wormlike micelles. In particular, we notice that the intensity of this instability (characterized by the fluctuating flow fields) increases as the Weissenberg number increases; however, beyond a critical value of it, this elastic instability and/or the flow field fluctuation is suppressed because of the breakage of long micelles. This is in contrast to the polymer solutions for which the flow field gradually transits to a more chaotic and turbulent-like state (or the so-called elastic turbulence state) as the Weissenberg number gradually increases. Additionally, we observe that the flow dynamics of these WLM solutions are strongly dependent on the type of micropore throat, the number of pore throats, and the spacing between two consecutive pore throats. An extensive discussion on the pressure drop and apparent viscosity is also presented in the present study. Although this study is carried out for a model porous media; however, we hope it will facilitate a better understanding of the flow behaviour of wormlike micellar solutions in an actual more complex porous media.
\end{abstract}

\maketitle

\section{\label{Intro}Introduction}
Surfactant molecules are amphiphilic, i.e., they contain both hydrophobic tail and hydrophilic head groups in their molecular architecture~\citep{myers2020surfactant,porter2013handbook}. As a result, when these molecules dissolve in a solvent like water, they tend to form an aggregate, called micelles. In particular, the hydrophobic tail groups are present inside a micelle, whereas the hydrophilic head groups are present in contact with the surrounding solvent~\citep{moroi1992micelles}. Once the concentration of the surfactant molecules exceeds a critical value, known as the critical micelle concentration (CMC), these micelles spontaneously self-assemble and form large aggregates of different shapes like spherical, ellipsoidal, wormlike, or lamellae~\citep{dreiss2007wormlike,dreiss2017wormlike}. A further increase in the surfactant concentration leads to the formation of an entangled network structure of these micelles, which ultimately originates a complex viscoelastic behaviour of the resulting solution. However, the rheological properties of these solutions, particularly of wormlike micellar (WLM) solutions, are found to be more complex than that seen for polymer solutions~\citep{anderson2006rheology,rothstein2003transient,rothstein2008strong,berret1997transient,walker2001rheology}. This is because this micelle network structure can undergo continuous scission and reformation mechanisms in a flow field, which is unlikely to happen for polymers due to the presence of strong covalent backbone in their molecular architecture. Due to their interesting physical, chemical, and rheological properties, these wormlike micellar solutions are widely used in many practical applications ranging from chemical, food, cosmetic to biomedical and many other industries~\citep{yang2002viscoelastic}.

In particular, wormlike micellar solutions, are routinely used in some processes, for instance, in the enhanced oil recovery (EOR)~\citep{wever2011polymers,raffa2016polymeric,mandal2015chemical} or in the groundwater remediation~\citep{mosler1996surfactants,dwarakanath1999anionic}. A micellar solution is forced to flow through a porous matrix under different flow and operating conditions in these applications. A typical porous matrix is composed of many interconnected micropores of various sizes and shapes. Despite these complex micropore structures, the macroscopic flow characteristics of a simple Newtonian fluid through such porous media like the relation between the pressure drop and velocity is well described by the famous Darcy's law~\citep{darcy1856fontaines}. However, for a complex fluid such as wormlike micellar solution or polymer solution, the flow physics can not be completely captured either by the Darcy's law or any other continuum law like the Blake-Kozeny-Carman~\citep{carman1997fluid} or the Ergun model~\citep{ergun1949fluid}. This is mainly due to the fact that, unlike a Newtonian fluid, the viscosity of a complex fluid would tend to depend on the local deformation rate, which again depends on the local shear or extensional flow fields (or mixed of both) present in the system. The shear-thinning and extensional hardening properties of a complex fluid (induced from the shear flow and extensional flow fields inside the porous matrix, respectively) and also the elasticity, greatly influence its apparent viscosity. This, in turn, substantially regulates the macroscopic behaviour like the pressure drop of a complex fluid during its flow through a porous media~\citep{sochi2010non}. In most of the investigations dealing with either micellar or polymer solutions, it has been found that after a critical value of the flow rate, the pressure drop across a porous media abruptly increased with the flow rate, and a highly non-linear trend between them was observed~\citep{marshall1967flow,deiber1981modeling,slattery1967flow,sobti2014creeping,tiu1997flow}.          

To understand the flow dynamics at the microscale in a porous media in detail, a model porous media is often considered in both experimental and numerical studies~\citep{browne2020pore}. For instance, a microchannel with many micropillars placed in it, is very often used to understand the complex flow features of viscoelastic fluids in a porous media. Using such model porous media, many complex and exciting flow behaviour of viscoelastic fluids were revealed. For instance, De et al.~\citep{de2017lane,de2018flow} found the formation of preferential paths or lanes during the flow of viscoelastic fluids in porous media. Furthermore, these paths were seen to be varied both spatially and temporally once the Weissenberg number gradually increased. Recently, Walkama et al.~\citep{walkama2020disorder} observed that the flow becomes chaotic and turbulent-like in such model porous media. Interestingly, they found that this chaotic behaviour is suppressed as some random geometric disorder is introduced into this model porous media. According to them, this is due to a gradual decrease in the stretching of the polymer microstructure with an increase in the porous media disorder, resulting from the formation of paths or lanes in the geometry. However, more recently, Haward et al.~\citep{haward2021stagnation} proposed that it is not always the disorder that can suppress the chaotic flow behavior of viscoelastic fluids, but it is the number of stagnation points (which are formed at the front and back surfaces of the micropillars where the maximum stretching of microstructure can happen due to the presence of strong velocity gradient) which ultimately controls this behaviour. To prove their hypothesis, they carried out experiments with a model porous media wherein the micropillars are initially arranged both in staggered and aligned configurations, and then they introduced random geometric disorder in both the geometries. They noticed that for the initial aligned arrangement, the introduction of geometric disorder actually increased the chaotic flow behaviour as opposed to that seen for the staggered configuration carried out by Walkama et al.~\citep{walkama2020disorder}.

Although a microchannel consisting of micropillars can mimic the flow characteristics of a real porous media; however, a more simplified geometry is composed of a microchannel with many micropore throats or constrictions present in it, can also serve the purpose~\citep{galindo2012microfluidic,browne2020pore}. Actually, the former model porous media consists of many such simplified geometries, which will allow a deeper understanding of the micro-scale flow physics. Many investigations have also been carried out in the literature using such a simplified geometry. For instance, recently, Browne et al.~\citep{browne2020bistability} performed an extensive experimental investigation on the flow characteristics of viscoelastic polymer solutions both in a single and an array of multiple pore constrictions. When the spacing between two pores is large, they found the formation of upstream eddies of similar character as that seen for an isolated pore throat. However, as the pore spacing decreases, they observed the existence of stochastically switching bistability flow states in the pores, namely, eddy-dominated state and eddy-free state. They proposed that this complex flow behaviour originates due to the stretching and relaxation of polymer molecules as they travel through the pore space. Ekanem et al.~\citep{ekanem2020signature} found the existence of the elastic turbulence for the flow of anionic hydrolyzed polyacrylamide (HPAM) polymer solution through a single micropore throat once the Weissenberg number exceeded a critical value. Furthermore, they also observed that the onset of this elastic turbulence depends on many factors such as the ionic strength, the type of cation in the anionic HPAM solution, and the nature of pore configuration. More recently, Kumar et al.~\citep{kumar2021numerical} did a numerical study  using the FENE-P viscoelastic constitutive equation and found the existence of multiple distinct and unstable flow structures inside the pores. They showed that the stretching of polymer molecules facilitates the eddy formation, whereas their relaxation suppresses its formation. Therefore, they proved the hypothesis of Browne et al.~\citep{browne2020bistability}, which states that these multiple flow states arise due to the stretching and relaxation of polymer molecules. Furthermore, De et al.~\citep{de2017viscoelastic} carried out a detailed numerical study based on the same viscoelastic constitutive equation, and found that the micropore throat structure, i.e., whether it is symmetric or asymmetric, has a strong influence on the flow characteristics. For instance, they observed that a symmetric micro pore throat generates more volume averaged normal stresses than that originated by an asymmetric one. Furthermore, the apparent viscosity was found to be more shear-thinning in the former geometry than that seen in the latter one.   

Therefore, from the aforementioned literature cited herein, it is seen that the flow dynamics of viscoelastic fluids in a model porous media like a channel with multiple pore throats or constrictions can become very complex depending upon the number of pore throats, their arrangements, flow conditions, fluid rheology, etc. Notably, almost all of these studies, either experimental or numerical, were carried out for polymer solutions. In contrast to this, there is no such detailed corresponding investigation available for the wormlike micellar solutions. One can expect more complexity in the flow dynamics of these solutions due to the presence of breakage and reformation mechanisms of these micelles. The importance of these two mechanisms was found to be very crucial in explaining many experimental observations pertained to wormlike micellar solutions but not seen in polymer solutions under otherwise identical conditions, for instance, the formation of lip-vortex in flows through a microfluidic cross-slot cell~\citep{kalb2018elastic,kalb2017role}, the unsteady motion past a translating sphere~\citep{sasmal2021unsteady}, flow dynamics in a long micro pore with step expansion and contraction~\citep{sasmal2020flow} and the flow dynamics in a Taylor-Couette cell~\citep{mohammadigoushki2019transient}, etc. All these studies were carried out based on the two-species Vasquez-Cook-McKinley (VCM) model for micelles~\citep{vasquez2007network}. This VCM model considers the wormlike micelles as an elastic segment composed of Hookean springs, which all together form an elastic network that can continuously break and reform. These breaking and reforming processes of this model are incorporated based on the discrete and simplified version of Cate's original theory for wormlike micelles~\citep{cates1987reptation}. This simplified model efficiently captures all the typical rheological characteristics of wormlike micellar solutions like the shear thinning, shear banding, extensional hardening and subsequent thinning, etc.~\citep{pipe2010wormlike,zhou2014wormlike,cromer2009extensional}, and this model can also be easily incorporated into a CFD platform to facilitate large scale continuum simulations for complex geometries with practical relevance. It also allows us to capture the
temporal and spatial variations in the number density of the long and short micelles. These breakage and reformation dynamics, as well as the variation in the number density of chain species, were missing in the earlier single species bead-spring models, for instance, Johnson-Segalman (JS) model~\citep{olmsted2000johnson,lu2000effects} or Bautista-Manero-Puig (BMP) model~\citep{bautista1999understanding}.
Hence, these models do not allow to relate the stress directly to the microstructural dynamics of the micellar molecules as does by the VCM model.  

This study aims to present a detailed investigation on the flow dynamics of wormlike micellar solutions through a model porous media consisting of a microchannel with single or multiple micropore throats present in it based on this two-species VCM model for micelles. In particular, we explore how the breakage rate of micelles, pore type, pore spacing, flow conditions, etc., would tend to influence the flow characteristics of micellar solutions in such model porous media. We hope it will ultimately facilitate a better understanding of the flow features of these complex fluids in a real and more complex porous media. The organization of the remaining part of this article is as follows: section~\ref{Problem} described the problem description and the governing equation, section~\ref{NumDetail} represent the details of the numerical solution procedure. In section~\ref{Results}, a detailed discussion of the results is presented, and finally, section~\ref{Con} summarizes the findings of this study.

\section{\label{Problem}Problem description and governing equations}
\begin{figure*}
    \centering
    \includegraphics[trim=0cm 5.4cm 0cm 0cm,clip,width=18cm]{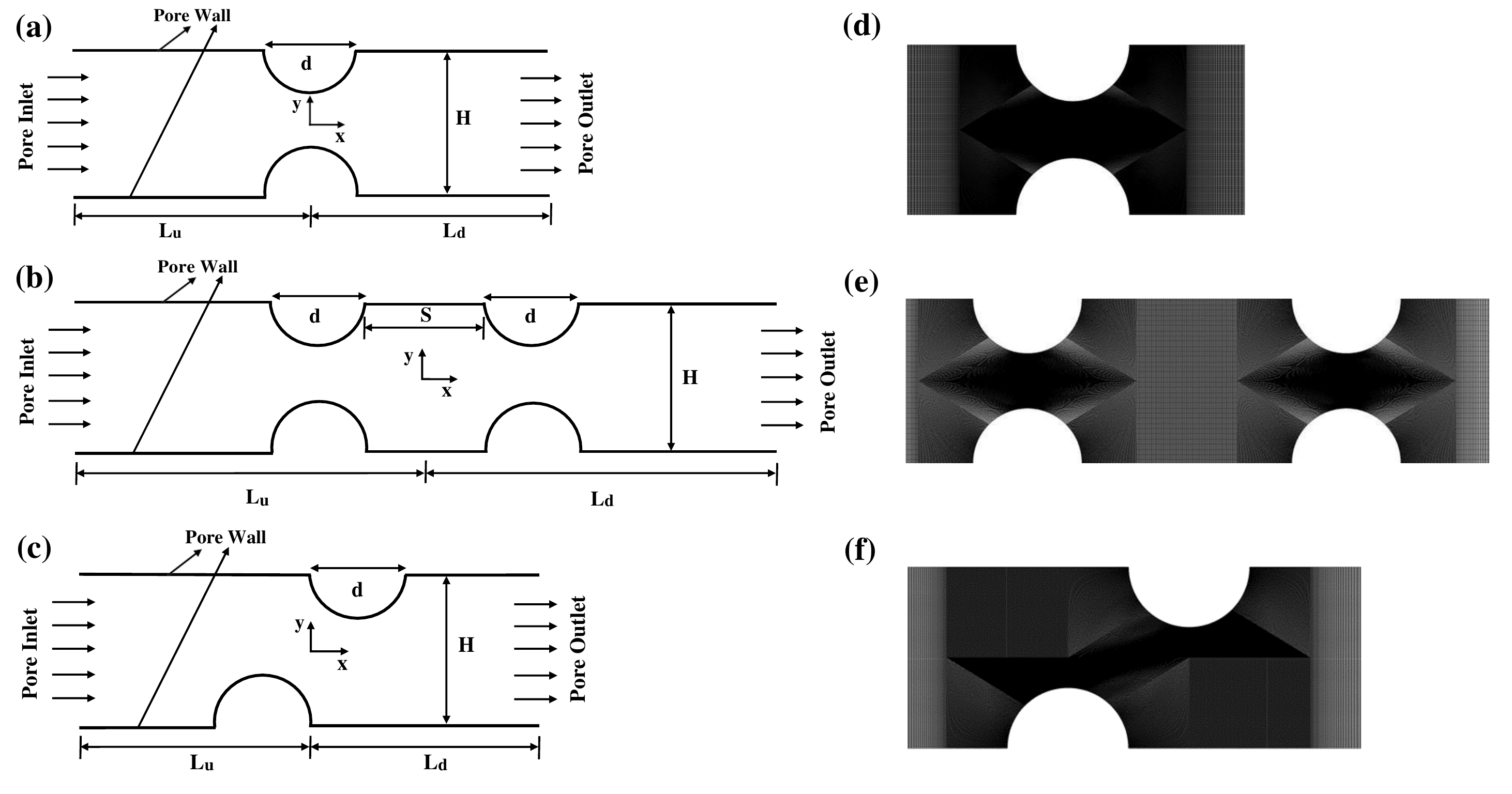}
    \caption{Schematic of the present problem considered in this study. Here sub-Figs.1 (a) and (b) represent the symmetric single and double micropore throats case, whereas sub-Fig.1 (c) represent the corresponding asymmetric single micropore throat case. Typical mesh densities used in the present study for single and double micropore throats for symmetric and asymmetric arrangements are shown in sub-Figs.1 (d-f). Here the flow direction is shown by arrows in the schematic.}
    \label{fig:1}
\end{figure*}

The present study aims to investigate the flow dynamics of a wormlike micellar solution through a microchannel with single and multiple micropore throats or constrictions present in it in the creeping flow regime, as shown schematically in Fig.~\ref{fig:1}. A pore throat is made of by placing two hemimicrocylinders of diameter $d$ in the microchannel of height $H$ with a fixed ratio of $\frac{d}{H} = 0.5$. In the case of double pore constrictions, the distance between the two hemimicrocylinders is $S$. We have simulated two arrangements of these micropores, namely, symmetric and asymmetric. Furthermore, for all the cases, the upstream $(L_{u})$ and downstream $(L_{d})$ length of the channel are kept as $25d$. This length was found to be sufficient so that it does not affect the flow dynamics around the micropore throat. Also, the micellar solution is assumed to be incompressible in nature.       
\begin{figure*}
    \centering
    \includegraphics[trim=0cm 0cm 0cm 0cm,clip,width=17cm]{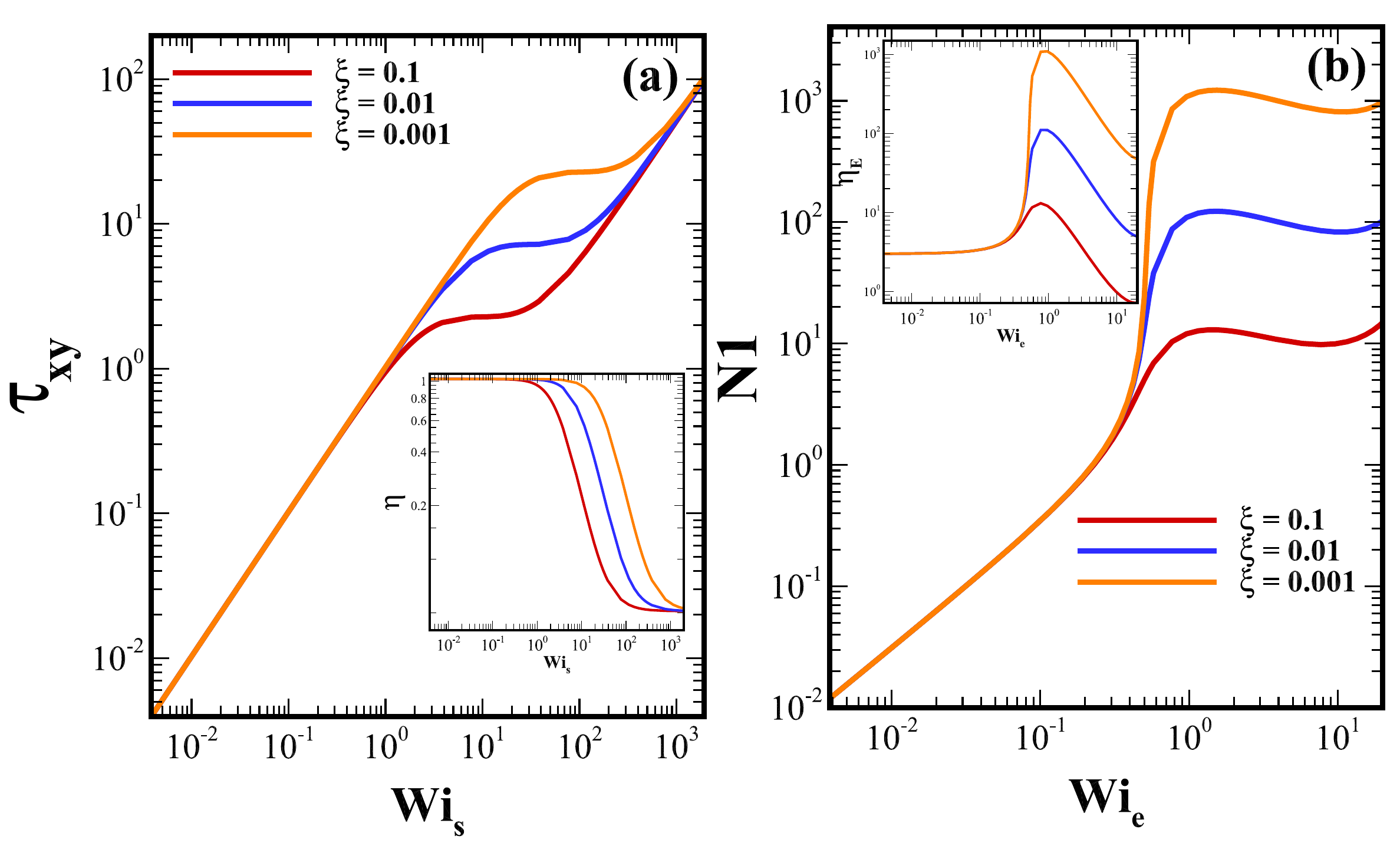}
    \caption{(a) Variation of the non-dimensional shear stress and shear viscosity (inset figure) with the non-dimensional shear rate (shear Weissenberg number, $Wi_{s}$) at different values of the micelle breakage rate (b) Variation of the non-dimensional first normal stress difference and extensional viscosity (inset figure) with the non-dimensional extensional rate (extensional Weissenberg number, $Wi_{e}$) at different values of the micelle breakage rate.}
    \label{fig:RheoFlow}
\end{figure*}
Under the above circumstances, the present flow field will be governed by the following equations, written in their dimensionless forms:\newline
Equation of continuity
\begin{equation}
\label{mass}
    \bm{\nabla} \cdot \bm{U} = 0
\end{equation}
Cauchy momentum equation 
\begin{equation}
\label{mom}
    -\nabla P + \nabla \cdot \bm{\tau} = 0
\end{equation}
In the above equations, $\bm{U}$, $t$ and $\bm{\tau}$ are the velocity vector, time and total extra stress tensor, respectively. In order to non-dimensionalize the above equations, all the spatial dimensions are scaled by the hemimicrocylinder diameter $d$, velocity is scaled by $d/\lambda_{eff}$, stress is scaled by the plateau modulus $G_0$ and time is scaled by $\lambda_{eff}$. Here $\lambda_{eff} = \frac{\lambda_{A}}{1+c_{Aeq}^{'}\lambda_{A}}$ is the effective relaxation time of the two-species VCM model, whereas $\lambda_{A}$ and $c_{Aeq}^{'}$ are the dimensional relaxation time and equilibrium breakage rate of the long worm A, respectively. It will be discussed in detail at the end of this section. In the present study, the Weissenberg and Reynolds numbers are defined as $Wi = \frac{\lambda_{eff}U_{in}}{d}$ and $Re = \frac{d U_{in} \rho}{\eta_{0}}$, respectively. A perfect inertialess (i.e., $Re = 0$) flow condition is simulated in this study by neglecting the convective terms of the Cauchy momentum equation. Here $\rho$ and $\eta_0$ are the density and zero-shear rate viscosity of the micellar solution, respectively. The total extra stress tensor, $\bm{\tau}$, for a wormlike micellar solution is given as:
\begin{equation}
    \label{totalStress}
    \bm{\tau} = \bm{\tau_{w}} + \bm{\tau_{s}}  
\end{equation}
where $\bm{\tau_{w}}$ is the non-Newtonian contribution from the wormlike micelles, whereas $\bm{\tau_{s}}$ is the contribution from that of the Newtonian solvent which is equal to $\beta \dot{\bm{\gamma}}$. Here the parameter $\beta$ is the ratio of the solvent viscosity to that of the zero-shear rate viscosity of the wormlike micellar solution and $\dot{\bm{\gamma}} = \nabla \bm{U} + \nabla \bm{U}^{T} $ is the strain-rate tensor. For the two-species VCM model, the total extra stress tensor is given by 
\begin{equation}
     \bm{\tau} = \bm{\tau}_{w} + \bm{\tau_{s}} = (\bm{A} + 2\bm{B})-\left(n_{A} + n_{B}\right)\bm{I} + \beta\dot{\bm{\gamma}}
\end{equation}
Here $n_{A}$ and $\bm{A}$ are the number density and conformation tensor of the long worm A respectively, whereas $n_{B}$ and $\bm{B}$ are to that of the short worm B. The temporal and spatial evaluation of the number density and conformation tensor for the short and long worms are written according to the VCM model as follows~\citep{vasquez2007network}:
\begin{equation}
    \label{nA}
    \mu\frac{Dn_{A}}{Dt} - 2\delta_{A} \nabla^{2}n_{A} = \frac{1}{2} c_{B} n_{B}^{2} - c_{A}n_{A}
\end{equation}
\begin{equation}
    \label{nB}
    \mu\frac{Dn_{B}}{Dt} - 2\delta_{B} \nabla^{2}n_{B} = - c_{B} n_{B}^{2} + 2 c_{A}n_{A}
\end{equation}
\begin{equation}
    \label{A}
    \mu \bm{A}_{(1)} + \bm{A} -n_{A} \bm{I} -\delta_{A} \nabla^{2}\bm{A} = c_{B} n_{B} \bm{B} - c_{A} \bm{A}
\end{equation}
\begin{equation}
    \label{B}
    \epsilon \mu \bm{B}_{(1)} + \bm{B} -\frac{n_{B}}{2} \bm{I} -\epsilon\delta_{B} \nabla^{2}\bm{B} = -2\epsilon c_{B} n_{B} \bm{B} + 2 \epsilon c_{A} \bm{A}
\end{equation}
Here the subscript $( )_{(1)}$ denotes the upper-convected derivative defined as $\frac{\partial()}{\partial t} + \bm{U}\cdot \nabla () - \left( (\nabla \bm{U})^{T} \cdot () + ()\cdot \nabla \bm{U}\right)$. The non-dimensional parameters $\mu$, $\epsilon$ and $\delta_{A,B}$ are defined as $\frac{\lambda_{A}}{\lambda_{eff}}$, $\frac{\lambda_{B}}{\lambda_{A}}$ and $\frac{\lambda_{A} D_{A,B}}{R^{2}}$, respectively. Here $\lambda_{B}$ is the relaxation time of the short worm $B$ and $D_{A, B}$ are the dimensional diffusivities of the long and short worms. Furthermore, according to the VCM model, the non-dimensional breakage rate $(c_{A})$ of the long worm A into two equally sized small worms B depends on the local state of the stress field, given by the expression $c_{A} = c_{Aeq} + \mu \frac{\xi}{3}\left( \dot{\bm{\gamma}}: \frac{\bm{A}}{n_{A}} \right)$. On the other hand, the reforming rate of the long worm A from the two short worms B is assumed to be constant, which is given by the equilibrium reforming rate, i.e., $c_{B} = c_{Beq}$. Here the non-linear VCM model parameter $\xi$ is the scission energy required to break a long worm into two equal-sized short worms. The significance of this parameter is that as its value decreases, the amount of stress needed to break a micelle increases. The values of the VCM model parameters chosen for the present study are as follows~\cite{khan2020effect,kalb2018elastic}: $\beta_{VCM} = 10^{-4}$, $\mu = 2.6$, $C_{Aeq} = 1.6$, $C_{Beq} = 0.8607$, $\epsilon = 0.005$, $\delta_{A} = \delta_{B}$ and $\xi = 0.001, 0.01, 0.1$. The behaviour of the present micellar solution with these VCM model parameters in standard rheological flows like shear and uniaxial extensional flows is shown in Fig.~\ref{fig:RheoFlow}. It can be clearly seen that the solution exhibits the shear-thinning property in shear flows and extensional hardening and subsequent thinning in uniaxial extensional flows, which are very often seen for a wormlike micellar solution. Furthermore, one can see that as the value of $\xi$ increases and/or the micelles are become easy to break, the shear-thinning tendency of the micellar solution increases, whereas the extensional hardening and subsequent thinning tendencies decrease.

\section{\label{NumDetail}Numerical details}
To solve the governing equations, namely, mass, momentum, viscoelastic constitutive, and number density evaluation equations mentioned in the previous section, the finite volume method based open-source computational fluid dynamics (CFD) code OpenFOAM~\cite{wellerOpenFOAM} and a recently developed rheoFoam solver available in rheotool~\cite{rheoTool} have been used in the present study. All the diffusion terms of these governing equations were discretized using the second-order accurate Gauss linear orthogonal interpolation scheme. All the gradient terms were discretized using the Gauss linear interpolation scheme. While the linear systems of the pressure and velocity fields were solved using the preconditioned conjugate solver (PCG) with DIC (Diagonal-based Incomplete Cholesky) preconditioner, the stress fields were solved using the preconditioned bi-conjugate gradient solver (PBiCG) solver with DILU (Diagonal-based Incomplete LU) preconditioner~\cite{ajiz1984robust,lee2003incomplete}. All the advective terms of the constitutive equations were discretized using the high-resolution CUBISTA (Convergent and Universally Bounded Interpolation Scheme for Treatment of Advection) scheme for its improved iterative convergence properties~\cite{alves2003convergent}. In the present study, the pressure-velocity coupling was established using the SIMPLE (Semi-Implicit Method for Pressure-Linked Equations) method, and the improved both side diffusion (iBSD) technique was used to stabilize the numerical solutions. The absolute tolerance level for the pressure, velocity, stress, and micellar concentration fields was set as $10^{-10}$.

A suitable grid with proper density and structure is also needed to get the final numerical results, which will ultimately be free from any numerical artifact. In the present study, it has been chosen by performing the standard grid independence study. In doing so, three different grid densities both for single and double pore throats, namely, G1, G2, and G3, consisting of different numbers of grid points on the throat surface as well as in the whole computational domain were created, and the simulations were run at the highest value of the Weissenberg number considered in the present study. After inspecting the results (in terms of the variation of the velocity, stress, and number densities of micelles at different probe locations in the computation domain) obtained for different grid densities, the grid G2 with a range of 149000-245650 (this number depends whether the channel contains single pore throat or multiple pore throats) hexahedral cells for the symmetric case and 180000-204100 hexahedral cells for the asymmetric case, were found to be adequate for the present study. During the making of any grid for any geometry, careful consideration is taken into account. For instance, a very fine mesh is created in the vicinity of the solid pore wall to capture the steep gradients of the velocity, stress, or concentration fields, whereas a relatively coarse mesh is created away from the solid wall, see sub-Figs.~\ref{fig:1}(d)-(f). Likewise, the grid independence study, a systematic time independence study, was also carried to choose an optimum time step size, and a non-dimensional time step size of 0.00001 was selected for all the cases. The computational domain and its meshing have been done with the help of the blockMeshDict subroutine available in OpenFOAM. Finally, appropriate boundary conditions were employed at different boundaries of the present computational domain to facilitate the simulations. On all solid surfaces, the standard no-slip and no-penetration boundary conditions for the velocity, i.e., $\bm{U} = 0$ were imposed, whereas a no-flux boundary condition was assumed for both the stress and micellar number density, i.e., $\textbf{n}\cdot \nabla \textbf{A} = 0$ and $\textbf{n}\cdot \nabla \textbf{B} = 0$ and $\textbf{n}\cdot \nabla {n_A} = 0$ and $\textbf{n}\cdot \nabla {n_B} = 0$, where $\textbf{n}$ is the outward unit normal vector. All the simulations were run in a parallel fashion with MPI (Message Passing Interface) interface facility available in OpenFOAM wherein each simulation was distributed among 8 to 12 CPU cores, each having 2 GB RAM. A detailed validation of the present numerical setup has already been presented in our earlier studies~\cite{sasmal2020flow,khan2020effect,khan2021elastic}, and hence it is not again performed here.  

\section{\label{Results}Results and discussion}
\subsection{Single micropore throat case}
\begin{figure}
    \centering
    \includegraphics[trim=0cm 0cm 2cm 0cm,clip,width=10cm]{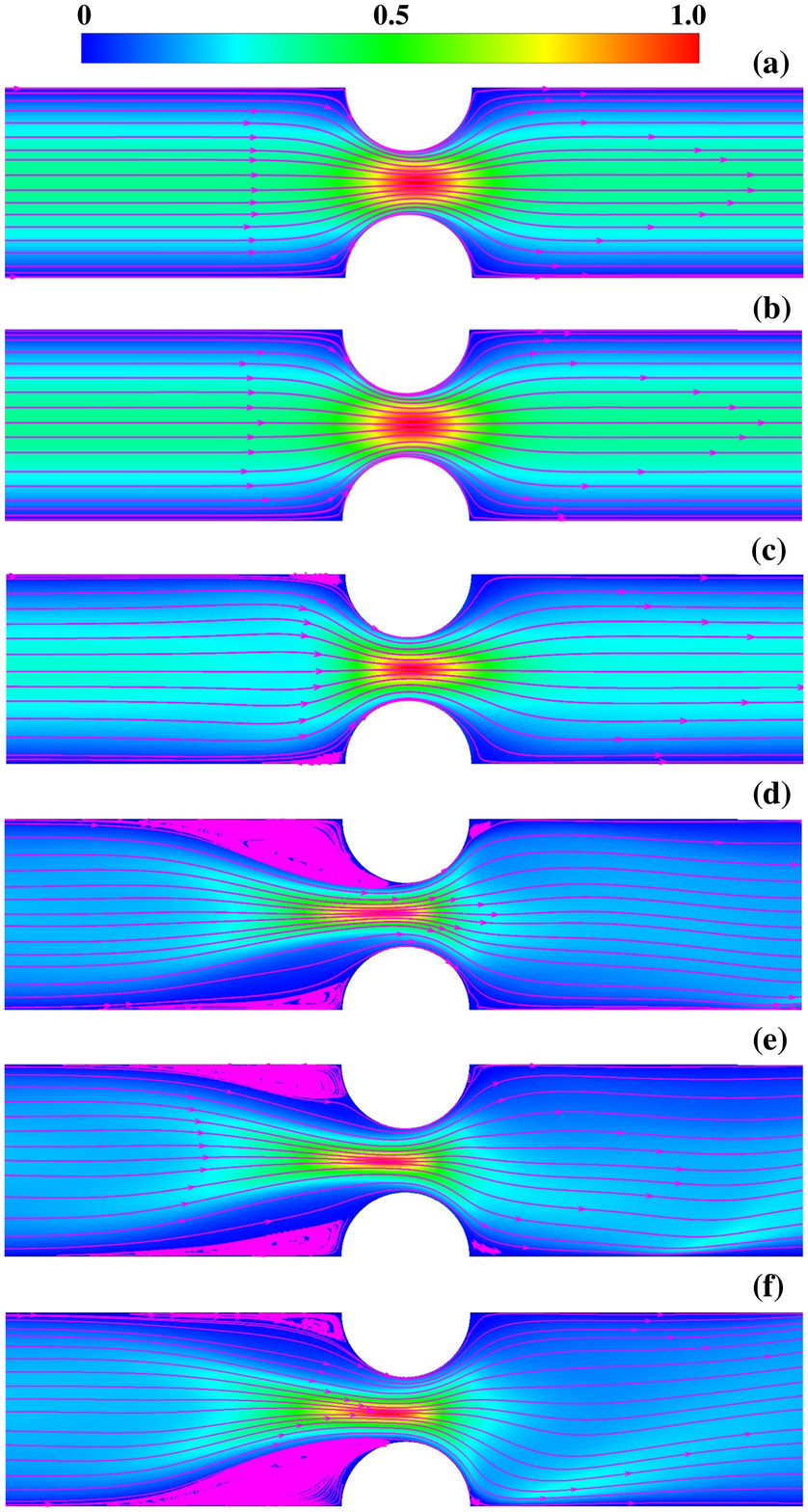}
    \caption{Representative streamline and velocity magnitude plots for a symmetric single micro pore throat case with $\xi = 0.01$. (a) Newtonian fluid (b) WLM, $Wi = 0.1$ (c) WLM, $Wi = 1.0$ and (d-f) WLM, $Wi = 3.0$ at three different times.}
    \label{fig:2}
\end{figure}

\begin{figure*}
    \centering
    \includegraphics[trim=0cm 4.5cm 0cm 2cm,clip,width=16cm]{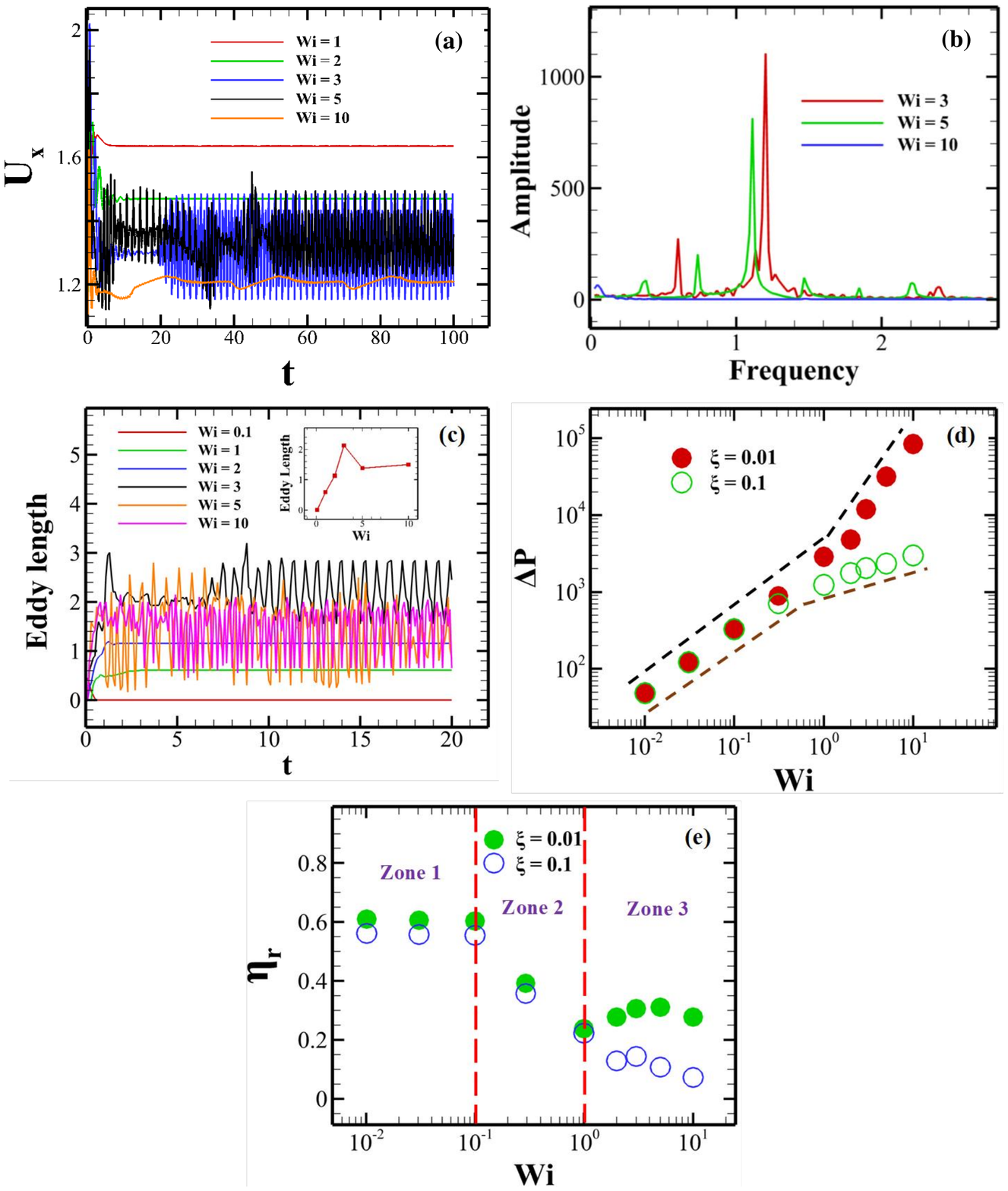}
    \caption{(a) Temporal variation of the non-dimensional stream-wise velocity component at a probe location placed downstream ($X = 1, Y = 0$) of the throat (b) The corresponding power spectral density plot of the velocity fluctuations at different Weissenberg numbers for a micelle breakage rate $\xi = 0.01$ (c) Temporal variation of the eddy length with the Weissenberg number (d) Variation of the non-dimensional pressure drop with the Weissenberg number and micelle breakage rate (e) Variation of the apparent viscosity with the Weissenberg number and micelle breakage rate. All the results are shown here for a symmetric single micropore throat case.}
    \label{fig:3}
\end{figure*}

First of all, the results for a single micropore throat are presented and discussed in this subsection. The streamlines, as well as velocity magnitude plots, are depicted in Fig.~\ref{fig:2} both for Newtonian and WLM solutions to investigate the velocity field inside the symmetric micropore throat in detail. For WLM solutions, three values of the Weissenberg number are chosen, namely, 0.1, 1, and 3, for a fixed value of the micelle breakage rate $\xi = 0.01$. As expected in the creeping flow regime,  the velocity field shows a perfect fore-aft symmetry for a Newtonian fluid, and there is no eddy present either upstream or downstream of the throat, as can be seen from sub-Fig.~\ref{fig:2}(a). A similar flow pattern is also observed for WLM solution at $Wi = 0.1$, sub-Fig.~\ref{fig:2}(b). This is due to the fact that at this low value of the Weissenberg number, the non-Newtonian characteristics of the WLM solution like the shear-thinning or elastic properties are expected to be very week, and hence it behaves like a Newtonian fluid. Furthermore, the flow remains in the steady-state, and a zone of high-velocity magnitude is formed at the center of the throat for both cases. However, as the Weissenberg number gradually increases, the flow becomes increasingly complicated for WLM solutions. For instance, at $Wi = 1$, two small eddies are seen to form both at the top and bottom upstream corners of the throat, sub-Fig.~\ref{fig:2}(c). Therefore, the fore-aft symmetry seen in the streamline profiles both for Newtonian and WLM solution with low Weissenberg number is destroyed at this value of the Weissenberg number. However, a symmetry along the horizontal plane passing through the center of the throat still exists, and the flow remains in a steady state. This is noticeable in sub-Fig.~\ref{fig:3}(a) wherein the temporal variation of the non-dimensional stream-wise velocity component is plotted at a probe location (X = 1.0, Y = 0) placed downstream of the throat . One can see that the velocity reaches a steady value with time at $Wi = 1$. 

As the Weissenberg number further increases to 3, the flow transits to an unsteady state, as can be evident from the variation in the streamline plots presented in sub-Figs.~\ref{fig:2}(d-f) at three different times as well as from the temporal variation of the stream-wise velocity shown in sub-Fig.~\ref{fig:3}(a). This unsteadiness is caused due to the emergence of elastic instability in the system. The presence of streamline curvature in the throat region and the generation of large tensile stresses downstream of the throat cause this elastic instability. The criteria for the onset of this purely elastic instability is developed by Pakdel and McKinley~\citep{pakdel1996elastic,mckinley1996rheological}, which is given as
\begin{equation}
    \left( \frac{\lambda_{eff}U}{\mathcal{R}} \frac{\tau_{xx}}{\eta \dot \gamma}\right) \ge M_{crit}^{2}
\end{equation}
In the above equation, $\mathcal{R}$ is the characteristic radius of the streamline curvature, $\tau_{xx}$ is the tensile stress in the flow direction, $\eta_{0}$ is the zero-shear-rate viscosity of the fluid, and $\dot{\gamma}$ is the characteristic value of the local deformation rate. Once the right-hand side parameter $M_{crit}$ exceeds a critical value, a purely elastic instability starts to appear. For instance, for the creeping flow of a constant shear viscosity viscoelastic fluid (or the so-called Boger fluid) past a cylinder confined in a channel, the value of $M_{crit}$ was found to be 6.2~\citep{mckinley1996rheological}, whereas for the flow in a lid-driven cavity it was 4.8~\citep{mckinley1996rheological,pakdel1998cavity}. A calculation of this parameter in a geometry similar to the present study was performed by Browne et al.~\citep{browne2020bistability}, and they found a range of values between 6 and 31 in their experiments depending on the dimension of the geometry and flow conditions. The unsteadiness observed due to this elastic instability is found to be aperiodic in nature at this Weissenberg number. This can be observed both in the plots of the temporal variation of the non-dimensional stream-wise velocity (at the same probe location as that presented for $Wi = 1.0$) shown in sub-Fig.~\ref{fig:3}(a) and in the power spectrum plot of the velocity fluctuations presented in sub-Fig.~\ref{fig:3}(b). In the latter plot, one can see that the velocity fluctuations are governed by a single dominant frequency with high amplitude, and also by some secondary frequencies with relatively low amplitude, thereby suggesting the presence of aperiodic unsteadiness in the flow field. Before the transition in this flow state, a regular periodic unsteadiness in the flow field was observed at the Weissenberg number in between 2 and 3 (the results are not shown here). Furthermore, at this Weissenberg number, the eddies are seen to form at the upper and lower corners of both upstream and downstream of the throat. However, the eddy size is seen to be always large upstream of the throat as compared to that seen downstream of it. Moreover, the eddy size is fluctuating with time. For instance, at $t = 19.1$, the eddy size is seen to be larger at the upstream top corner than that seen at the bottom one, sub-Fig.~\ref{fig:2}(d). At a later time $t = 19.5$, an opposite trend is seen, as can be evident from sub-Fig.~\ref{fig:2}(f). In between these two times, for instance, at $t = 19.3$, the eddy size is found to be similar at both the locations, see sub-Fig.~\ref{fig:2}(e). The temporal fluctuation in the eddy size can also be seen in sub-Fig.~\ref{fig:3}(c) wherein the top upstream eddy size is plotted against the time. Likewise, the stream-wise velocity, it also shows aperiodic fluctuation. This kind of periodic displacement of the eddies between the top and bottom upstream corners was also observed in the experiments dealing with polymer solutions~\citep{browne2020bistability,ekanem2020signature} as well as in simulations~\citep{kumar2021numerical} with similar geometry as considered in the present study (see the supplementary video 1).

\begin{figure}
    \centering
    \includegraphics[trim=3.2cm 11.5cm 0cm 0cm,clip,width=10.5cm]{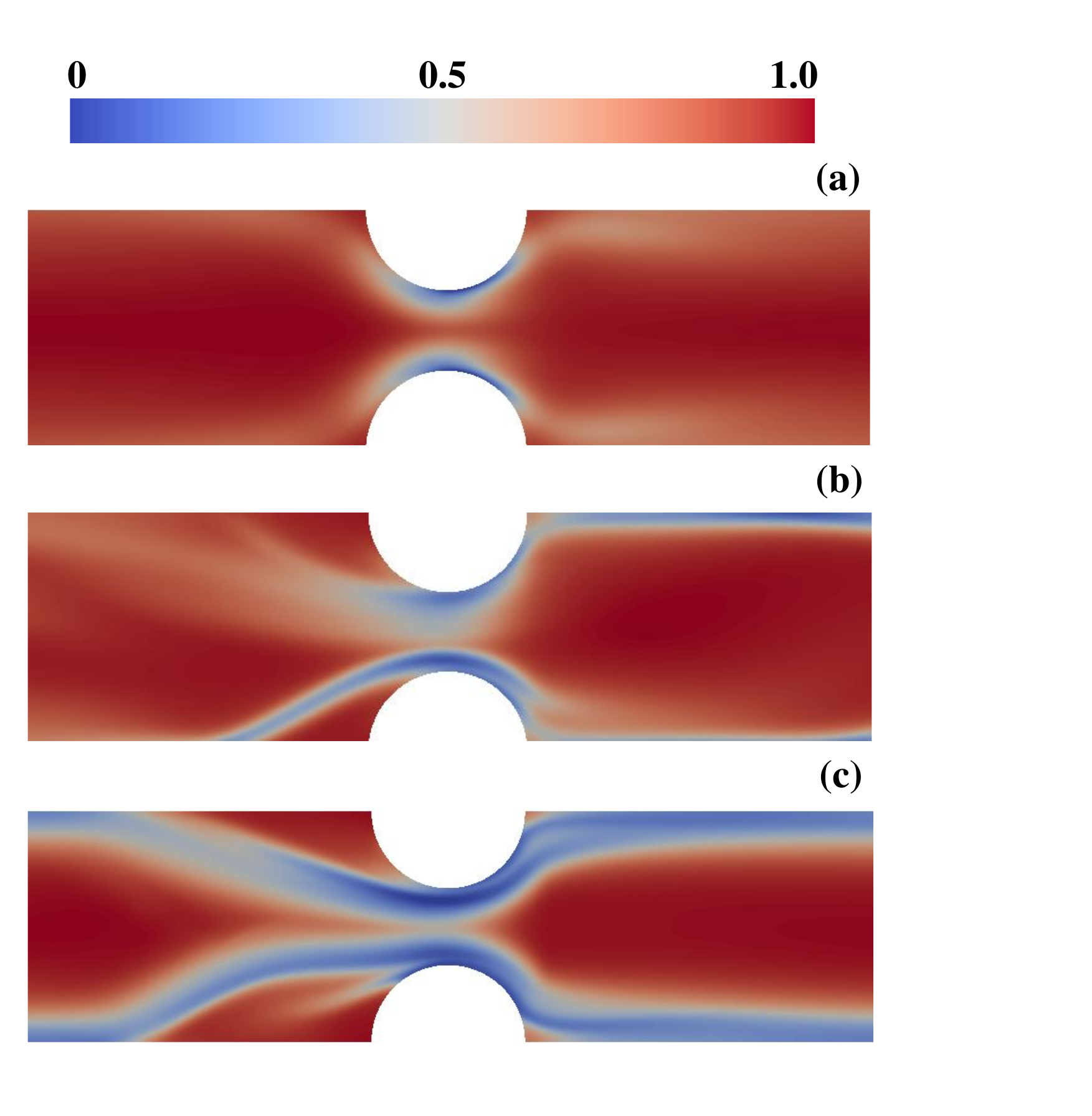}
    \caption{Distribution of the long chain number density of a WLM solution in the symmetric micropore throat at $\xi = 0.01$ (a) $Wi = 1.0$ (b) $5$ and (c) $10$.}
    \label{fig:6}
\end{figure}

As the Weissenberg number further increases to 5, once again, one can see the instability in the flow from both the temporal variation of the stream-wise velocity component (sub-Fig.~\ref{fig:3}(a)) and the eddy length presented in sub-Fig.~\ref{fig:3}(c). However, at this Weissenberg number, both the amplitude and frequency of fluctuations decrease (sub-Fig.~\ref{fig:3}(b)). The unsteadiness is still governed by a dominant frequency; however, some other secondary frequencies are also seen to present likewise, it was seen at $Wi = 3$. As the Weissenberg number is further incremented to 10, the velocity field shows less fluctuations as compared to that seen either at $Wi = 3$ or 5. This can be again seen from both the temporal variation of the stream-wise velocity and power spectral density plot presented in sub-Fig.~\ref{fig:3}(a) and (b), respectively. This is stark contrast to that observed for a polymer solution~\citep{browne2020bistability,kumar2021numerical,ekanem2020signature}. For a polymer solution, the velocity field gradually transits from steady to unsteady periodic, then quasi-periodic, and finally, irregular and chaotic turbulent-like state (known as the elastic turbulence) as the Weissenberg number gradually increases. This is mainly due to the increase in the elastic forces with the Weissenberg number. However, for a WLM solution, the reason for not showing more fluctuating behaviour with the gradual increment in the Weissenberg number can be explained as follows: as the Weissenberg number gradually increases and/or the velocity magnitude increases (here, the Weissenberg is incremented by increasing the inlet velocity in the pore, i.e., $U_{in}$), both the shear and extensional flow strengths increase within the geometry. This results in the gradual increase of the breakage of long micelles into smaller ones. This can be evident in Fig.~\ref{fig:6} wherein the number density of long micelles is depicted at three different Weissenberg numbers, namely, 1, 3, and 10. One can see that at $Wi = 10$ (sub-Fig.~\ref{fig:6}(c)), the number density of long micelles becomes very low as compared to that seen either at $Wi = 3$ (sub-Fig.~\ref{fig:6}(a)) or 5
(sub-Fig.~\ref{fig:6}(b)). Due to this breakage of long micelles, the elastic stresses generated in the system decrease as it will be in the extensional thinning region under this flow condition, see Fig.~\ref{fig:RheoFlow}. This, in turn, inhibits the unsteadiness in the flow field at high Weissenberg numbers. This kind of transition in the flow state in a WLM solution was also experimentally observed by  Haward et al.~\citep{haward2019flow} wherein they studied the flow behaviour past a microcylinder confined in a channel.

\begin{figure*}
    \centering
    \includegraphics[trim=0cm 10cm 0cm 0cm,clip,width=16cm]{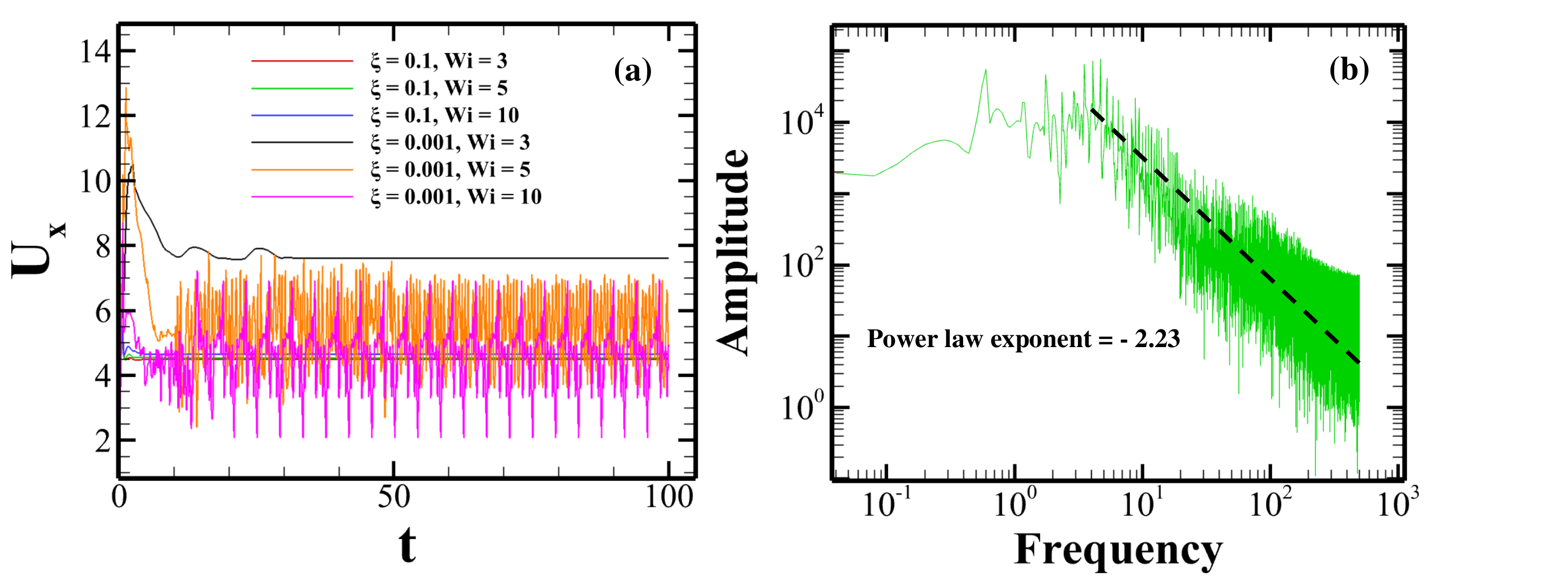}
    \caption{(a) Effect of the micelle breakage rate and Weissenberg number on the temporal variation of the stream-wise velocity component in a symmetric micropore throat (b) The corresponding power spectral density plot of the velocity fluctuations at $\xi = 0.001$ and $Wi = 5$.}
    \label{fig:4}
\end{figure*}

Therefore, one can expect that the micelle breakage rate (which is controlled by the non-linear parameter $\xi$ in the present VCM model) would have a strong influence on the flow dynamics of WLM solutions in the micropore throat. To explicitly show this, we have carried out simulations for two other values of the micelle breakage rate, namely, $\xi = 0.1$ and 0.001. Note that as the value of $\xi$ increases, the long micelles are become progressively easier to break, as mentioned earlier. At $\xi = 0.1$, it can be seen that the stream-wise velocity (at the same probe location as that presented for $\xi = 0.01$) reaches a constant value with time at both Weissenberg numbers (3 and 5), thereby suggesting that the flow is in the steady-state, for instance, see Fig.~\ref{fig:4}(a).
There are two reasons for this: first of all, at this value of $\xi$, the shear-thinning property of the WLM solutions is more pronounced as compared to that seen at the other two relatively lower values of $\xi$ (as can be seen from sub-Fig.~\ref{fig:RheoFlow}(a)). This property of a viscoelastic fluid has a tendency to suppress the elastic instability~\citep{casanellas2016stabilizing}. On the other hand, at this value of $\xi$, the extensional thickening property of the WLM solutions is seen to be less as compared to that seen for other values of the micelle breakage rate, see sub-Fig.~\ref{fig:RheoFlow}(b). This is simply due to the easy breaking of the micelles at this value of $\xi$. Unlike the shear-thinning property, the extensional thickening tendency of a viscoelastic fluid promotes elastic instability as well as elastic turbulence. Therefore, the flow field remains in the steady-state at this value of $\xi$ due to the increase in the shear-thinning and decrease in the extensional thickening tendencies of the micellar solution. 

\begin{figure}
    \centering
    \includegraphics[trim=0.8cm 12cm 0cm 0cm,clip,width=9cm]{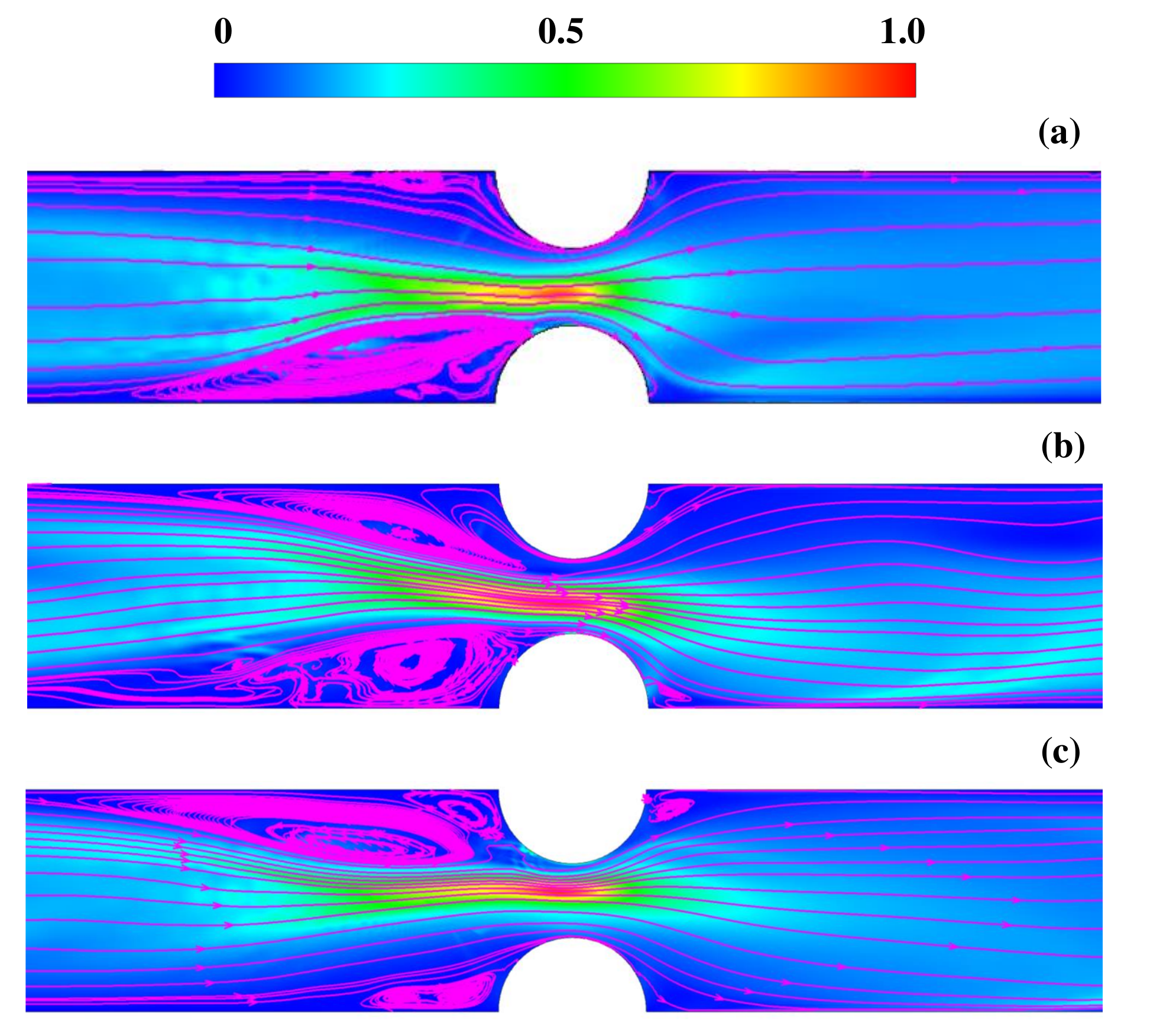}
    \caption{(a-c) Representative streamline and velocity magnitude plots for the symmetric  micropore throat at $\xi = 0.001$ and $Wi = 5$ at three different times.}
    \label{fig:5}
\end{figure}

On the other hand, the corresponding streamlines and velocity magnitude plots at the value of $\xi = 0.001$ are presented in Fig.~\ref{fig:5} at $Wi = 5$.
Once again, the eddies formed at the top and bottom upstream corners of the throat show temporal fluctuations in their size as that observed at $\xi = 0.01$ (Fig.~\ref{fig:2}). Moreover, the eddies are now broken into smaller secondary eddies at some instances, for instance, see the results at $t = 21.9$ in sub-Fig~\ref{fig:5}(c). Also, the shape of the eddies is now more distorted than that seen at $\xi = 0.01$. Therefore, the flow now becomes more chaotic and turbulent-like as compared to that seen at $\xi = 0.01$. This is more evident in sub-Fig.~\ref{fig:4}(a) wherein more fluctuations can be seen in the temporal variation of the stream-wise velocity at the same probe location as that presented for $\xi = 0.01$ or 0.1. The corresponding power spectrum of the velocity fluctuations is depicted in sub-Fig.~\ref{fig:4}(b). It can be seen that the velocity fluctuations are governed by more than one dominant frequencies with a broad range of values. At higher frequencies, it shows a power-law decay with an exponent of -2.23, which indicates a broad range of temporal and spatial scale fluctuations in the velocity. An almost similar value of the power-law exponent was also observed in the pressure fluctuations in a similar geometry dealing with polymer solutions~\citep{ekanem2020signature}. Note that this exponent value is higher than the value for the Kolmogorov turbulence, which has a power-law decay of | 5/3 |~\citep{kolmogorov1941degeneration}. All these suggest the signature of the elastic turbulence present in the system at this combination of the micelle breakage rate and flow conditions, which is driven by the elastic stresses but not the inertia~\citep{groisman2000elastic,steinberg2021elastic}.        

The variation of the non-dimensional pressure drop with the Weissenberg number across the micropore throat is shown in sub-Fig.~\ref{fig:3}(d) for two values of the micelle breakage rate, namely, 0.01 and 0.1. Irrespective of the value of the micelle breakage rate, the pressure drop always increases with the Weissenberg number. This is due to the fact that as per the Darcy's law for the flow through porous media, the pressure drop and the flow rate are directly proportional to each other
\begin{equation}
    U = -\frac{K}{\mu \epsilon} \frac{\Delta P}{L}
\end{equation}
Note that here the flow rate and/or the mean velocity $(U)$ is related to the Weissenberg number, and the latter is changed by changing the former one. In the above equation, $K$ and $\epsilon$ are the permeability and porosity of the porous medium, respectively. On the other hand, $\mu$ is the viscosity that is constant for a Newtonian fluid, but for the present WLM solution, it depends on the flow conditions. For the present flow, the pressure drop will be contributed from two parts, namely, due to the shear-thinning behaviour $(\Delta P)_{shear}$ and due to the presence of elastic stresses $(\Delta P)_{elastic}$. For constant values of $K$ and $\epsilon$, the former effect will tend to reduce the pressure drop as the viscosity will tend to decrease, whereas the latter effect will tend to increase it due to the increase in the viscosity contributed by the elastic stresses. For the micelle breakage rate $\xi = 0.1$, the pressure drop varies non-linearly with the Weissenberg number. It continues to increase up to a critical value of the Weissenberg number of around 1, and then it starts to decrease. This is because of the fact the at this value of $\xi$, the WLM solution shows a considerable extent of shear-thinning behaviour after this critical value of the Weissenberg number, which actually overcomes the influence of the elastic stresses on the pressure drop. As a result, overall, the pressure drop starts to decrease. On the other hand, for the micelle breakage rate $\xi = 0.01$, once again,  a highly non-linear pattern is observed. However, at this value of the micelle breakage rate, the pressure drop starts to increase rapidly after this critical value of the Weissenberg number. This is because of the increase in the pressure loss due to the increase in the elastic stresses at this micelle breakage rate, which ultimately leads to the emergence of elastic instability. This sudden rise of the pressure drop has also been seen in flows of viscoelastic fluids for other geometries, like the flow past an obstacle~\citep{varshney2017elastic,grilli2013transition} or the flow through a serpentine microchannel~\citep{yang2020experimental,li2017efficient} wherein the elastic instability and elastic turbulence phenomena have been observed. 

The ratio of the pressure drop of wormlike micellar solutions to that of a Newtonian fluid under otherwise identical conditions, defined as the apparent viscosity, is given as
\begin{equation}
    \eta_{app} = \frac{\Delta P_{WLM}}{\Delta P_{Newt}}
\end{equation}
The variation of this apparent viscosity with the Weissenberg number is shown in sub-Fig.~\ref{fig:3}(e) again for two values of the micelle breakage rate, namely, 0.01 and 0.1. There are three distinct regimes present in this figure. In the first regime, the apparent viscosity shows a plateau region as in this low Weissenberg regime, the WLM solution behaves like a Newtonian fluid. As the Weissenberg number is gradually incremented, a second regime, known as the so-called shear-thinning regime, is observed wherein the apparent viscosity decreases with $Wi$. As the Weissenberg number further increases and exceeds a critical value, a third regime is seen wherein the apparent viscosity starts to increase with $Wi$. This regime is called the shear-thickening regime, wherein an unsteadiness in the flow field and/or the presence of elastic instability is observed. This trend in the variation of the apparent viscosity with the Weissenberg was also found in experiments dealing with polymer solutions~\citep{ekanem2020signature}. However, for the present case of a WLM solution, the apparent viscosity again starts to decrease once the Weissenberg number exceeds another critical value. This is due to the breakage of long micelles into smaller ones, which ultimately leads to the reduction of fluctuations in the velocity field, as discussed earlier. The presence of these three distinct regimes is clearly evident for the micelle breakage rate $\xi = 0.01$, whereas for $\xi = 0.1$, mostly two regimes, namely, the plateau and shear-thinning regimes, are seen to present due to the absence of the elastic instability at this value of the micelle breakage rate.

\begin{figure}
    \centering
    \includegraphics[trim=0cm 0cm 15cm 0cm,clip,width=8cm]{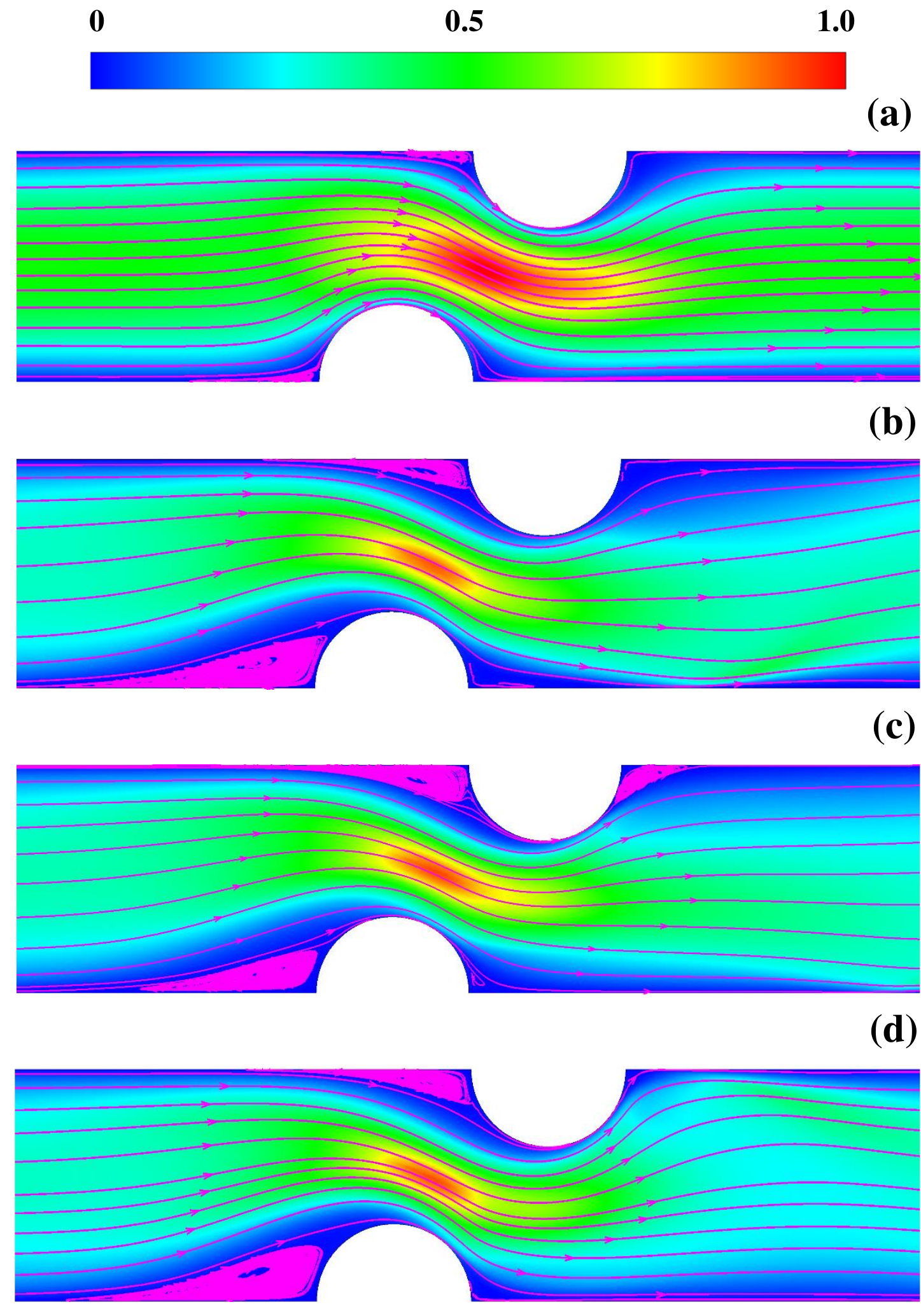}
    \caption{Representative streamline and velocity magnitude plots for an asymmetric micropore throat at $\xi = 0.01$ (a) $Wi = 1.0$ and (b-d) $Wi = 3.0$ at three different times.}
    \label{fig:7}
\end{figure}

\begin{figure*}
    \centering
    \includegraphics[trim=6cm 0cm 6cm 0cm,clip,width=16cm]{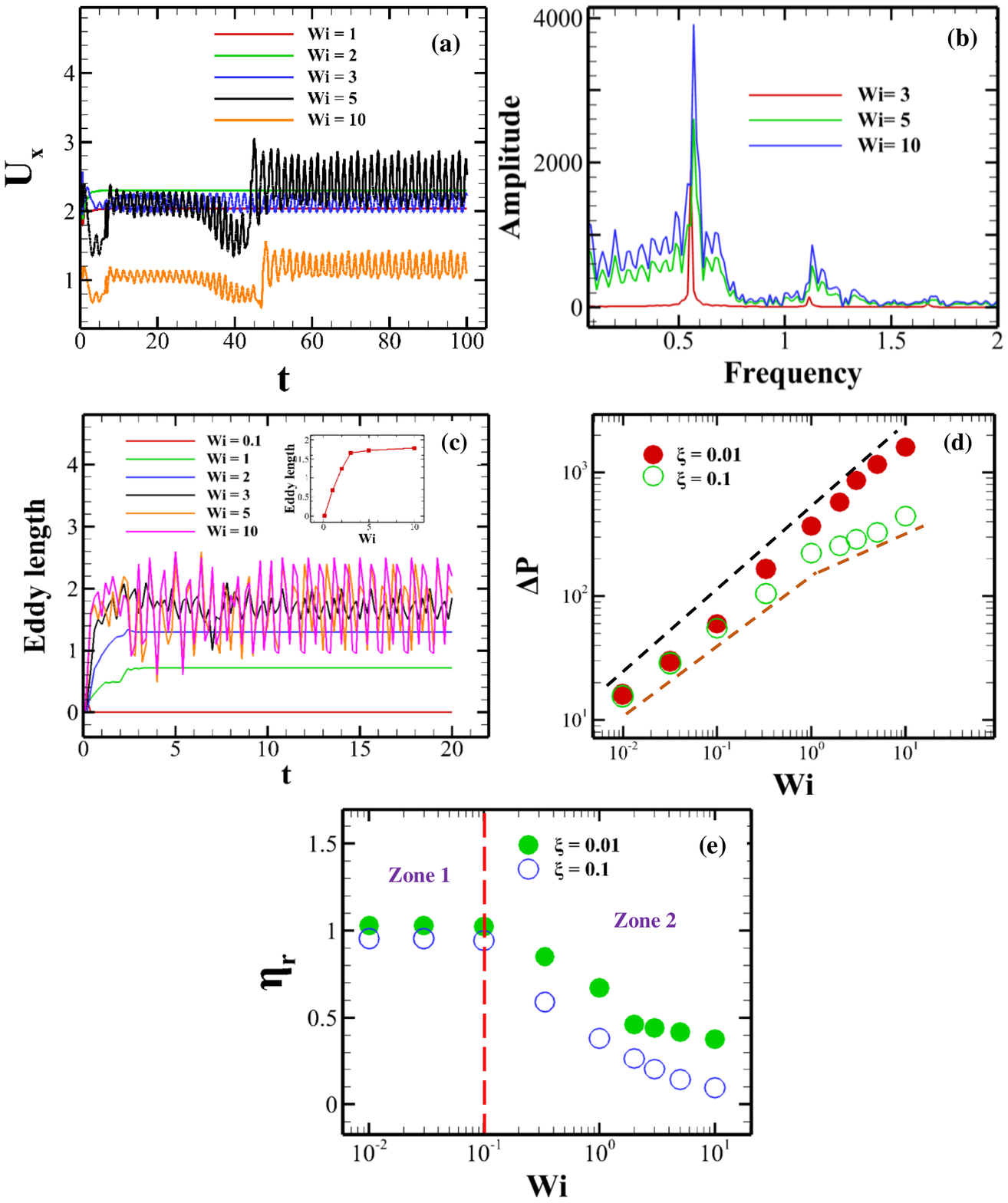}
    \caption{(a) Temporal variation of the stream-wise velocity component at a probe location for an asymmetric micropore throat (b) The corresponding power spectral density plot of the velocity fluctuations at different Weissenberg numbers $\xi = 0.01$ (c) Temporal variation of the top upstream corner eddy length for an asymmetric micropore (d) Variation of the pressure drop with the Weissenberg number and micelle breakage rate (e) Variation of the apparent viscosity with the Weissenberg number and micelle breakage rate.}
    \label{fig:8}
\end{figure*}

Finally, we show how the micropore structure can also influence the flow dynamics inside it. To do so, we have carried out simulations in an asymmetric micropore throat, as schematically shown in sub-Fig.~\ref{fig:1}(c).  Figure~\ref{fig:7} shows the streamlines and velocity magnitude plots
at different Weissenberg numbers for a micelle breakage rate of $\xi = 0.01$. At $Wi = 1$, the flow is steady and two small eddies are formed in the upstream corners of the throat. As the Weissenberg gradually increases, likewise the symmetric micropore throat, the flow also becomes unsteady after a critical value of the Weissenberg number, for instance, see the results presented at $Wi = 3$ for three different times in sub-Figs.~\ref{fig:7}(b-d). At this $Wi$, the eddy size becomes large as compared to that seen at $Wi = 1$ which again fluctuates with time (see the supplementary video 2). This can be seen in sub-Fig.~\ref{fig:8}(c) 
wherein the temporal variation of the eddy size is presented. Furthermore, one can see that the time-averaged eddy size gradually increases with the Weissenberg number, for instance, see the inset figure in the same sub-Fig. This is in contrast to that seen in the case of a symmetric micropore throat for which a non-monotonic dependence of the time-averaged eddy size on the Weissenberg number is seen (see the inset figure in sub-Fig.~\ref{fig:3}(c)). This is because of the transition of the flow field from an unsteady towards a steady one at high Weissenberg numbers due to the breakage of long micelles in the latter case.

\begin{figure}
    \centering
    \includegraphics[trim=2cm 10.5cm 0cm 0cm,clip,width=10cm]{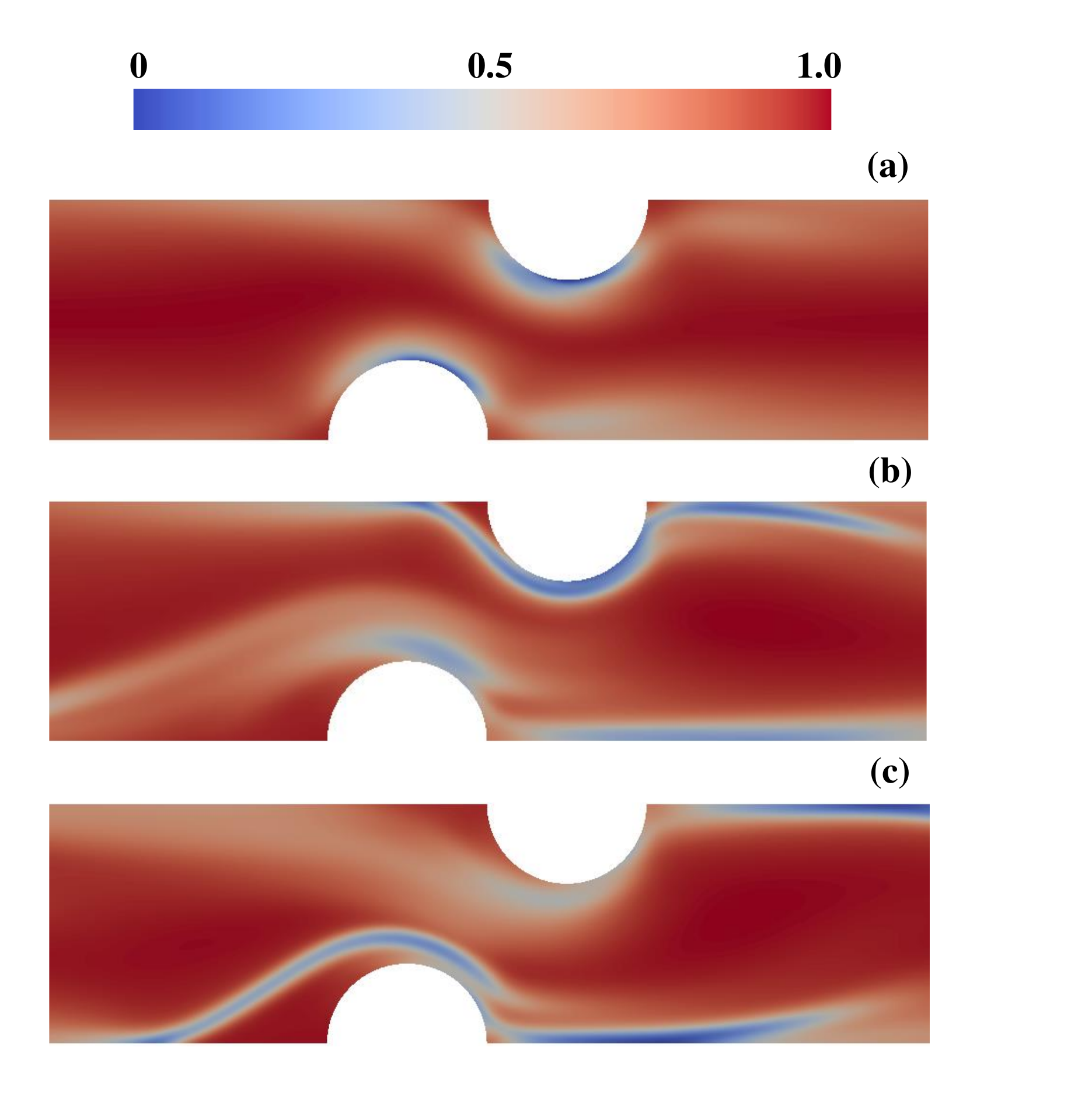}
    \caption{Distribution of the long chain number density of a WLM solution in an asymmetric micro pore throat at $\xi = 0.01$ (a) $Wi = 1.0$ (b) $Wi = 5$ and (c) $Wi = 10$.}
    \label{fig:9}
\end{figure}

In the case of flow through an asymmetric micro pore throat, the flow gradually transits from steady to periodic and then quasi-periodic as the Weissenberg number gradually increases. This can be evident from the temporal variation of the stream-wise velocity presented in sub-Fig.~\ref{fig:8}(a) as well as from the power spectral density plot presented in sub-Fig.~\ref{fig:8}(b). From the latter plot, one can see that the velocity fluctuations are governed by a single dominant frequency at $Wi = 3$, thereby suggesting the existence of a regular periodic flow state at this $Wi$. At higher Weissenberg numbers, for instance, at Wi = 5 and 10, the velocity fluctuations are also characterized by some secondary frequencies along with a primary dominant frequency, and hence the flow state is said to be in the quasi-periodic state. Furthermore, it can be seen that the amplitude of the power spectrum increases with the Weisseneberg number, thereby indicating that the intensity of the velocity fluctuations increases with the Weissenberg number. However, within the ranges of conditions encompassed in the present study, the flow field inside an asymmetric pore throat does not transit from a quasi-periodic to steady one at high Weissenebrg numbers likewise it was seen for a symmetric pore throat. This can be explained from the distribution of the number density of long micelles for this pore geometry presented in Fig.~\ref{fig:9}. 
It can be seen that the breakage of long micelles is always more in the case of symmetric pore throat than that seen for an asymmetric one under otherwise identical conditions, particularly see the results at $Wi = 10$ (sub-Fig.~\ref{fig:6}(c) for symmetric and sub-Fig.~\ref{fig:9}(c) for asymmetric pore throat). The reason for this is that the area for the flow of fluid in a symmetric micro pore throat is less as compared to that available for an asymmetric pore. Therefore, the magnitude of the velocity gradient in the vicinity of the throat is larger for the former pore than that for the latter one. This leads to more breakage of long micelles in a symmetric pore throat than that in an asymmetric pore throat. Due to the presence of a large flow area in an asymmetric pore throat, the fluid can pass easily through the throat area.     

The corresponding variation of the pressure drop and apparent viscosity with the Weissenberg number and micelle breakage rate for the flow through an asymmetric micropore throat is presented in sub-Figs.~\ref{fig:8}(d) and (e), respectively. Once again, the pressure drop increases with the Weissenberg number irrespective of the values of the micelle breakage rate, as it was seen for a symmetric micropore throat. However, there are some differences present in the trend. For instance, at $\xi = 0.01$, an almost linear dependence is observed in the variation of the pressure drop for an asymmetric micropore throat, whereas a non-linear dependence was seen for a symmetric micropore throat. This is because of the fact the at this value of $\xi$, the WLM solution shows a considerable extent of shear-thinning behaviour, whereas it also shows a reasonable extent of elastic behaviour as the Weissenberg number gradually increases. Therefore, there is a competitive influence present between these two oppositely acting effects on the pressure drop, which ultimately nullifies each other effects at this value of the micelle breakage rate and pore geometry. However, for the micelle breakage rate $\xi = 0.1$, similar behaviour is observed as that seen for a symmetric pore throat. The corresponding variation of the apparent viscosity with the Weissenberg number and micelle breakage rate is shown in sub-Fig.~\ref{fig:8}(e). For both the micelle breakage rates, the presence of two regimes can be seen for this pore geometry, namely, the plateau regime at low Weissenberg numbers and the shear-thinning regime at high Weissenberg numbers. However, there is no shear-thickening regime present for this pore geometry as it was seen for a symmetric pore throat. Therefore, one can say that the symmetric micro pore throat is more prone to the generation of elastic instability and elastic turbulence than that for an asymmetric one under otherwise identical conditions. 
\begin{figure*}
    \centering
    \includegraphics[trim=0cm 0cm 0cm 0cm,clip,width=17.5cm]{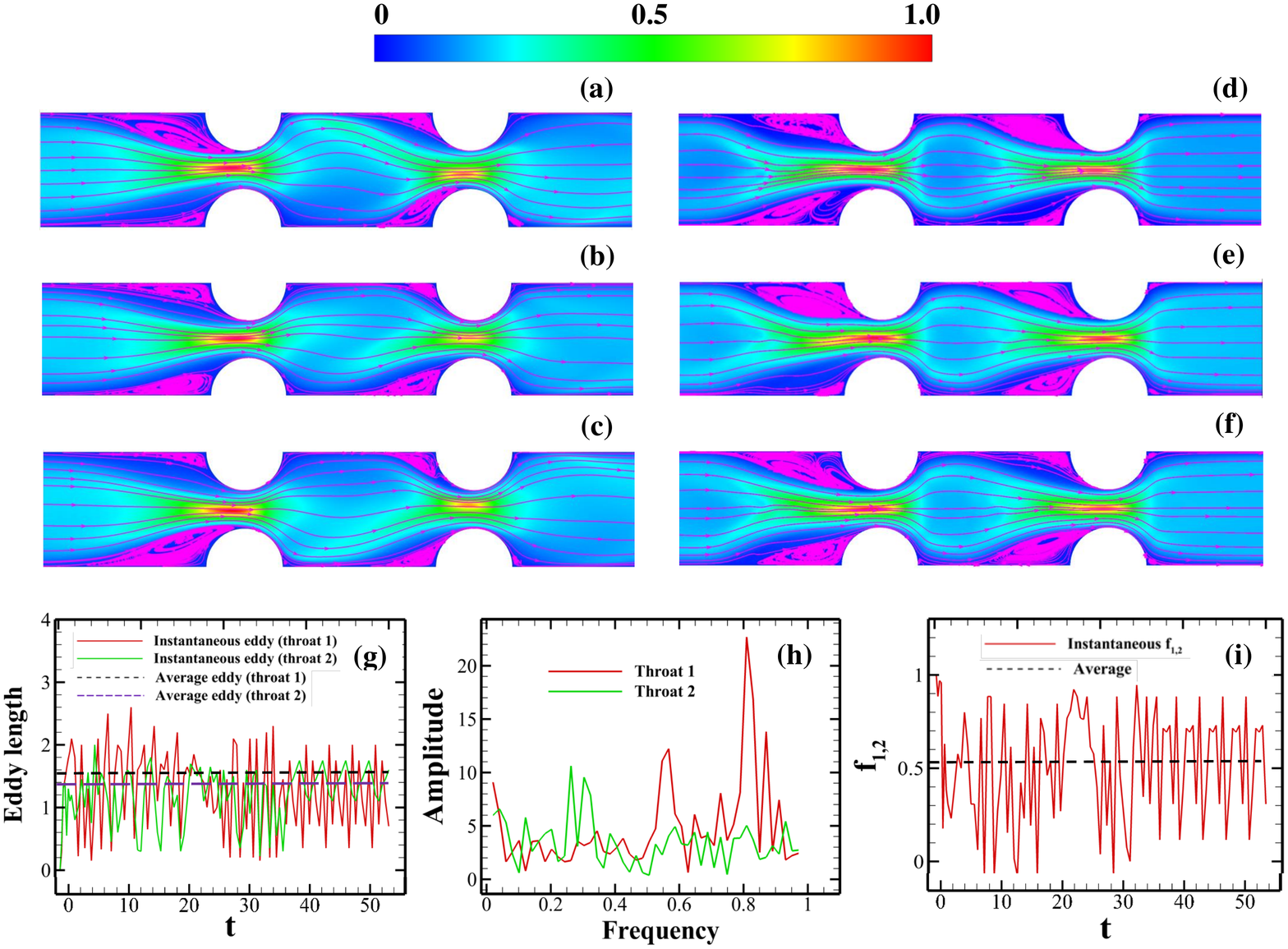}
    \caption{Representative streamline and velocity magnitude plots for symmetric double micro pore throats with a spacing of $S = 2$  at $\xi = 0.01$ (a-c) $Wi = 3$  (d-f) $Wi = 10$ at three different times (g) The temporal variation of the eddy length and its time-averaged value at throat 1 and throat 2 at $Wi = 5$ and $\xi = 0.01$ (h) The corresponding power spectral density plot of the eddy length fluctuation (i) The temporal variation of the correlation function $f_{1,2}$ and its time-averaged value at $Wi = 5$ and $\xi = 0.01$.}
    \label{fig:10}
\end{figure*}
\begin{figure*}
    \centering
    \includegraphics[trim=0cm 0cm 0cm 0cm,clip,width=17.5cm]{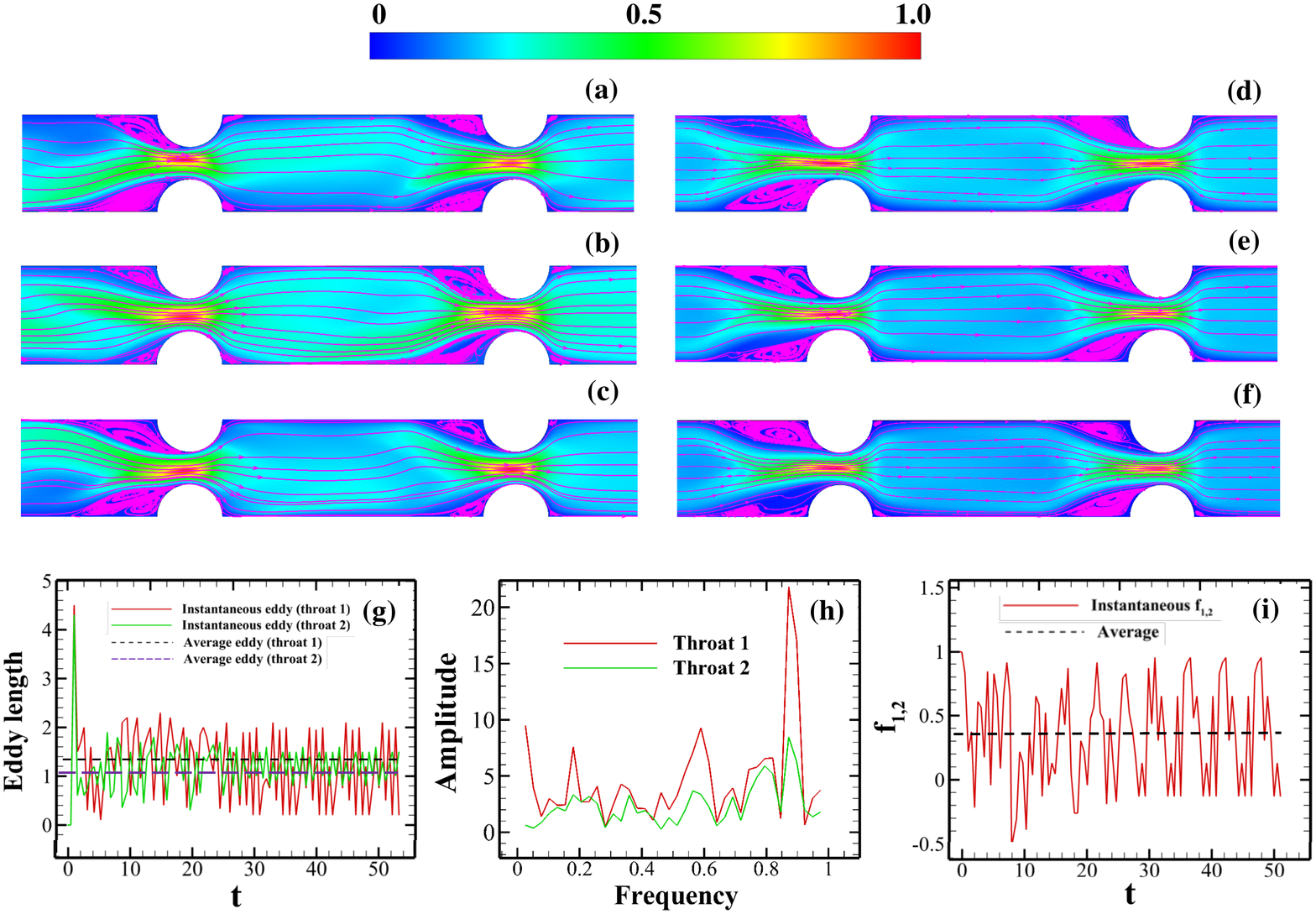}
    \caption{Representative streamline and velocity magnitude plots for symmetric double micro pore throats with a spacing of $S = 5$  at $\xi = 0.01$ (a-c) $Wi = 3$  (d-f) $Wi = 10$ at three different times (g) The temporal variation of the eddy length and its time-averaged value at throat 1 and throat 2 at $Wi = 5$ and $\xi = 0.01$ (h) The corresponding power spectral density plot of the eddy length fluctuation (i) The temporal variation of the correlation function $f_{1,2}$ and its time-averaged value at $Wi = 5$ and $\xi = 0.01$.}
    \label{fig:11}
\end{figure*}
\begin{figure*}
    \centering
    \includegraphics[trim=0cm 0cm 2cm 0cm,clip,width=14cm]{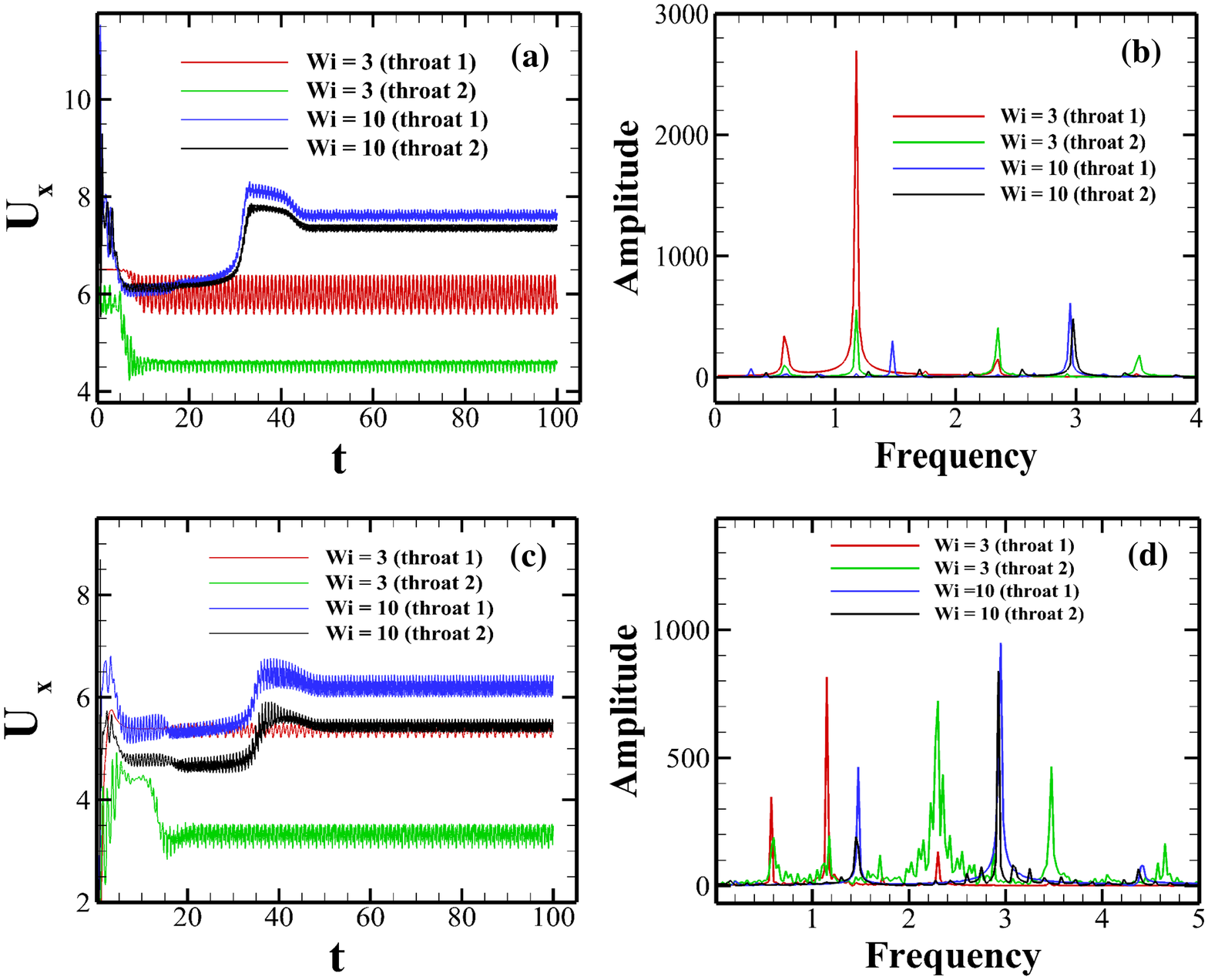}
    \caption{(a) Temporal variation of the stream-wise velocity component at a probe location placed at the middle in between throat 1 and 2 with a spacing $S = 2$ at $\xi = 0.01$ and (b) the corresponding power spectral density plot of the velocity fluctuations. Sub-Figs.(c) and (d) present the corresponding results for the spacing $S = 5$.}
    \label{fig:12}
\end{figure*}
\begin{figure}
    \centering
    \includegraphics[trim=0cm 17cm 0cm 0cm,clip,width=9cm]{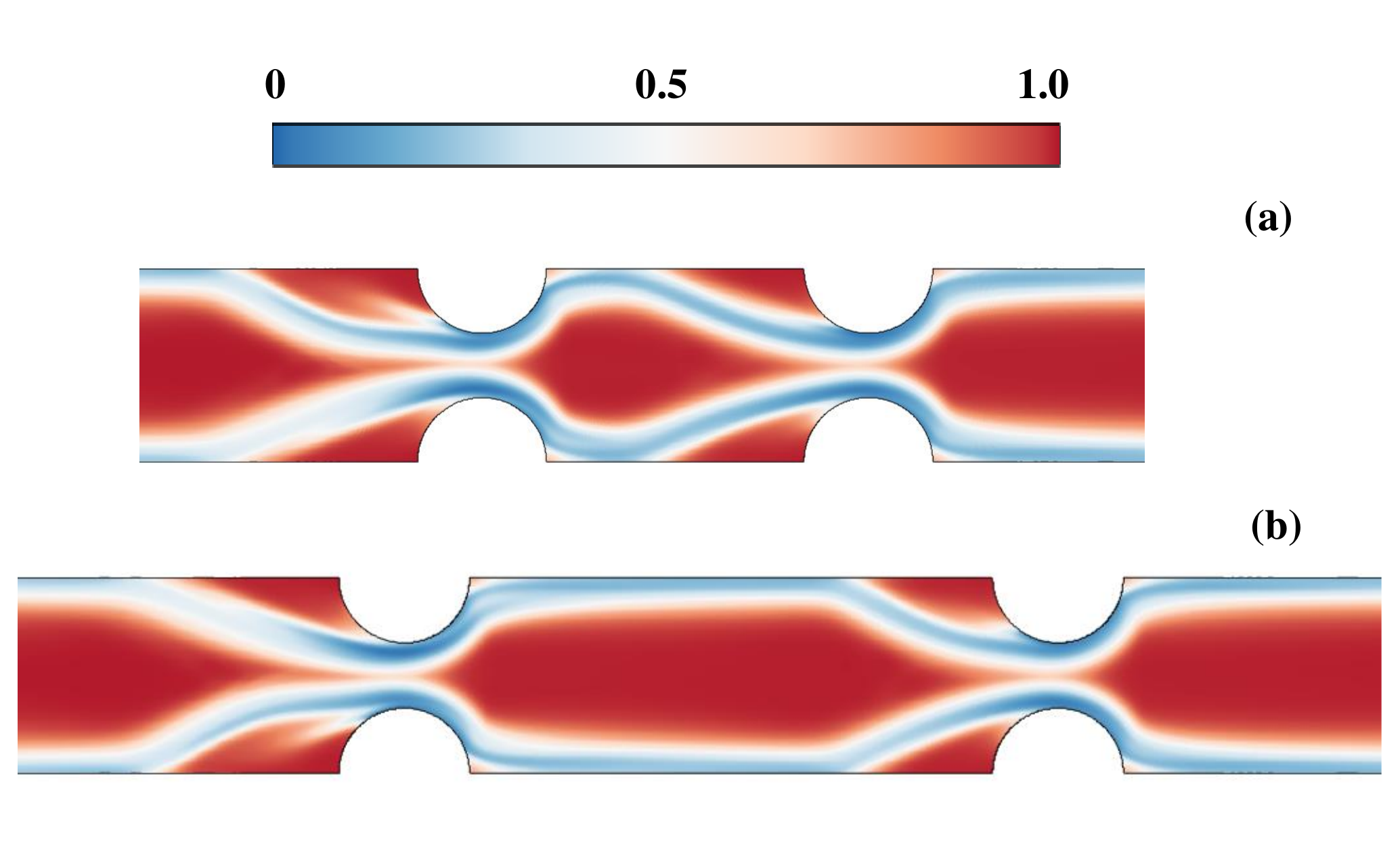}
    \caption{Distribution of the long chain number density of a WLM solution for a symmetric double micro pore throat at $\xi = 0.01$ and $Wi = 10$ (a) $S = 2$ and (b) $S = 5$.}
    \label{fig:13}
\end{figure}
\begin{figure*}
    \centering
    \includegraphics[trim=0cm 0cm 0cm 0cm,clip,width=14cm]{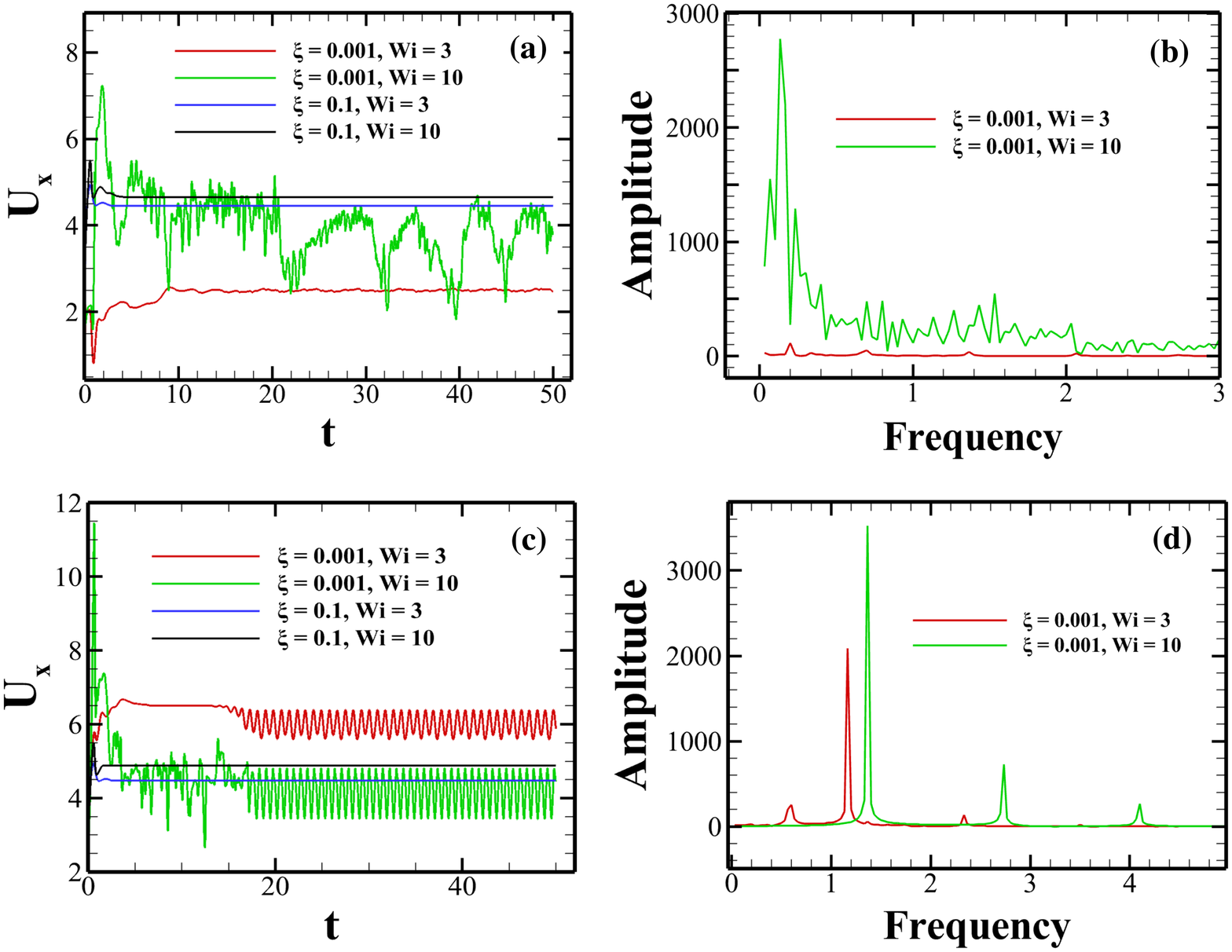}
    \caption{(a) Temporal variation of the stream-wise velocity component at a probe location placed at the middle in between throat 1 and throat 2 for two different values of $\xi$, namely, $0.001$ and $0.1$ (b) The corresponding power spectral density plot of the velocity fluctuations. Sub-Figs.(c) and (d) present the corresponding results for the spacing $S = 5$.}
    \label{fig:14}
\end{figure*}
\subsection{Multiple micro pore throats case}
While the single pore case provides many insights into the flow physics, it is the multi pores case that needs to be analyzed as it is seen in most practical applications. However, the single pore case would obviously facilitate a better understanding of the physics associated with the case of multi pores. Therefore, we have carried out simulations with two micropore throats with different pore spacing (namely, $S$ = 2 and 5, see sub-Fig.~\ref{fig:1}(b)) and five micropore throats with a pore spacing of 0.5 between two consecutive pores. Figure~\ref{fig:10} shows the streamlines and velocity magnitude plots for double micropore throats with a spacing of 2 at two Weissenberg numbers, namely, 3 (sub-Fig.~\ref{fig:10}(a-c)) and 10 (sub-Fig.~\ref{fig:10}(d-f)) for a micelle breakage rate $\xi = 0.01$. Likewise, the single pore throat, in this case of double pore throats, the eddies are also formed at both upstream top and bottom corners of both the first and second throats. Once again, as the Weissenberg number exceeds a critical value, the size of the eddies becomes fluctuating in nature, which is evident in the variation of the streamline profiles presented at three different times as well as in their temporal variation of length presented in sub-Fig.~\ref{fig:10}(g) (see the supplementary video 3). However, it can be seen that the eddies at the second throat are smaller in size than that seen at the first throat. It is more obvious if one can see the time-averaged value shown in the same sub-Fig.~\ref{fig:10}(g) presented at $Wi = 5$. This is primarily due to the fact that the micelles tend to relax downstream of the first throat, and hence they will have less influence on the vortex formation upstream of the second throat. A similar kind of observation was also seen in the flows of polymer solutions through two micro constrictions~\citep{kumar2021numerical}. The corresponding power spectral density plot of the eddy length fluctuations for both the first and second throats is presented in sub-Fig.~\ref{fig:10}(h). One can easily see that both the magnitude and frequency of eddy length fluctuations are higher for the first throat than that seen for the second one.  In order to establish an instantaneous correlation between the eddy length of two throats, a correlation function $f_{1,2}$ is defined as below~\citep{kumar2021numerical}
\begin{equation}
      f_{1,2} = 1 - \frac{2|L_{e,1} - L_{e,2}|}{max(L_{e,1}, L_{e,2})}
\end{equation}
where $L_{e,1}$ and $L_{e,2}$ are the eddy lengths for the first and second throats, respectively. A value of $f_{1,2} \rightarrow 1$ corresponds to the situation of similar eddies formed at throats 1 and 2, whereas $f_{1,2} \rightarrow -1$ indicates a maximum difference between them. The temporal variation of this correlation function as well as its time-averaged value are shown in sub-Fig.~\ref{fig:10}(i). The time-averaged value of the correlation function is found to be slightly more than 0.5, and hence one can say that the eddies formed at the first and second throats are partially correlated with each other.

As the spacing between the two throats increases, say to 5, a similar kind of trend is seen in the streamline profiles, eddy length, PSD plot of the eddy length fluctuations, etc., as those seen in the case of throat spacing 2, see Fig.~\ref{fig:11}. However, there are certain quantitative differences present in the results. For instance, the eddy length decreases as the throat spacing increases, as seen from sub-Fig.~\ref{fig:11}(g). This is particularly more evident for the eddy formed upstream of the second throat. The value of the correlation function $f_{1,2}$ also decreases as the throat spacing increases as can be seen from sub-Fig.~\ref{fig:11}(i). This can be explained with the help of the local Deborah number, which can be defined as $De = \frac{\lambda_{eff} U}{S}$ where $S$ is the distance between the two throats. As this distance increases, the Deborah number decreases, and hence the influence of the fluid elasticity on the flow characteristics also decreases, particularly around the second throat. This, in turn, results in a decrease of the correlation function between the eddy length of the two throats.

The temporal variation of the stream-wise velocity and the corresponding PSD plot of the velocity fluctuations are shown in Fig.~\ref{fig:12} for two values of the pore throat spacing, namely, 2 (sub-Figs.~\ref{fig:12}(a) and (b)) and 5 (sub-Figs.~\ref{fig:12}(c) and (d)) at two values of the Weissenberg number, namely, 3 and 10. It is seen that the velocity at the first pore throat is always larger than that seen at the second throat. However, the difference between them decreases as the Weissenberg number increases or the pore spacing decreases. The velocity at both throats shows fluctuations that are quasi-periodic in nature, as can be evident from the PSD plot presented in sub-Figs.~\ref{fig:12}(b) and (d). For throat spacing 5, both the magnitude and frequency of velocity fluctuations increase with the Weissenberg number, see sub-Fig.~\ref{fig:12}(d). In contrast to this, for throat spacing 2, the magnitude of velocity fluctuations decreases with the Weissenberg number, as one can see from sub-Fig.~\ref{fig:12}(d). This can be explained by the breakage and reformation of micelles which occur during the flow of WLM solution from one throat to another. To do so, the distribution of the number density of long micelles in shown in Fig.~\ref{fig:13} for two throat spacings, namely, 2 (sub-Fig.~\ref{fig:13}(a)) and 5 (sub-Fig.~\ref{fig:13}(b)) at a fixed Weissenberg number of 10. One can clearly see that the breakage of long micelles into short micelles is more for throat spacing 2 than that seen for 5, particularly near the second throat. Therefore, it results in lowering the elastic effects, and hence the velocity fluctuations.   

To show the effect of the micelle breakage rate on the flow dynamics for the two pore throats case explicitly, we have carried out additional simulations for other two values of the micelle breakage rate, namely, 0.001 and 0.1. The non-dimensional stream-wise velocity and the corresponding power spectral density of the velocity fluctuations at the middle of the two throats are plotted against the time for two throat spacings, namely, 2 (sub-Figs.~\ref{fig:14}(a) and (b)) and 5 (sub-Figs.~\ref{fig:14}(c) and (d)). It can be seen that for the micelle breakage rate $\xi = 0.1$, the flow fields remain in the steady-state irrespective of the throat spacing; likewise, it was seen for the single micropore throat case. This is again due to the easy breakage of micelles at this of $\xi$. However, for the micelle breakage rate $\xi = 0.001$, a fluctuating flow field is seen to present for both throat spacings. For throat spacing 5, a quasi-periodic unsteadiness in the flow field is seen, see sub-Fig.~\ref{fig:14}(d), whereas for throat spacing 2, a more chaotic and turbulent-like flow structure is observed, which can be evident from the PSD plot presented in sub-Fig.~\ref{fig:14}(b). This is because of the increase in the extensional hardening properties of the micelles (see Fig~\ref{fig:RheoFlow}) at this value of $\xi$, which thereby leading to the onset of this more chaotic flow state. Finally, we have carried out simulations for geometry with five micro pore throats, and the flow dynamics inside it is found to be similar in nature as that seen for the single and double micropore throats case, at least as far as the influence of the micelle breakage rate is concerned. Once again, the flow field becomes fluctuating in nature as the Weissenberg number exceeds a critical value (see the supplementary video 4). All the spaces between two successive pore throats are filled up by the eddies whose size and shape fluctuate with time. However, we have not found out the existence of a state where no eddy is present in the pore space at any instant of time; likewise, it was observed in the case of polymer solutions~\citep{browne2020bistability,ekanem2020signature}. The fluctuations are always more near the first pore throat, which are then slowly transferred towards the rest of the pore throats.

\section{\label{Con} Conclusions}
This study presents an extensive numerical investigation of the flow dynamics of wormlike micellar solutions (WLMs) through a model porous media consisting of a microchannel with single or multiple micropore throats present in it. A two-species Vasquez-Cook-Mckinley (VCM) viscoelastic constitutive model for the wormlike micelles has been used to realize the rheological behaviour of the present WLM solutions. As the Weissenberg number exceeds a critical value, a purely elastic instability is emerged in the flow field. This kind of instability, purely driven by the elastic forces, was already seen in many recent experiments and simulations dealing with polymer solutions in a similar kind of geometry~\citep{browne2020bistability,ekanem2020signature,kumar2021numerical}. In polymer solutions, the flow field transits from steady to quasi-periodic and then chaotic and turbulent-like structure (the so-called elastic turbulence) as the Weissenberg number gradually increases. In contrast to this, for the present case of wormlike micellar solutions, the transition from a quasi-periodic state to more chaotic and turbulent-like flow state with the gradual increment in the Weissenberg number may get inhibited due to the breakage of long micelles. This kind of flow behaviour for the WLMs was also seen experimentally in the case of flow past a microcylinder~\citep{haward2019flow}. Furthermore, the onset of such elastic instability is greatly influenced by the micropore throat arrangement. For instance, the intensity of the flow field fluctuations, originated due to this elastic instability, is found to be more for a symmetric micropore throat than that seen for an asymmetric one. The signature of this elastic instability is marked by an increase both in the pressure drop across the throat and the apparent viscosity defined by the ratio of the pressure drop of WLMs to that of Newtonian fluid under otherwise identical flow conditions. Moreover, the variation of both these pressure drop and apparent viscosity depends on the Weissenberg number, micelle breakage rate and micropore throat arrangement.

The flow dynamics of WLMs can also be greatly influenced by the presence of another or multiple micropore throats inside the microchannel, and also by the spacing between two successive micropore throats. As the spacing between the two micropore throats increases, the correlation in the flow dynamics decreases due to a decrease in the local Deborah number. The intensity of the flow field fluctuations, arising due to the elastic instability, is seen to be always more in the first throat, which is then slowly transferred to the rest of the throats. All in all, we hope this study will facilitate a better understanding of the flow dynamics of wormlike micellar solutions through a more complex and real porous media, which has widespread applications ranging from the enhanced oil recovery to ground water remediation.

\section{Acknowledgements}
The authors would like to thank IIT Ropar for providing the funding through the ISIRD research grant (Establishment 1/2018/IITRPR/921) to carry out this work. 

\section{Availability of data}
The data that supports the findings of this study are available within the article.
\nocite{*}

\section{References}
\bibliography{aipsamp}

\begin{thebibliography}{67}%
\makeatletter
\providecommand \@ifxundefined [1]{%
 \@ifx{#1\undefined}
}%
\providecommand \@ifnum [1]{%
 \ifnum #1\expandafter \@firstoftwo
 \else \expandafter \@secondoftwo
 \fi
}%
\providecommand \@ifx [1]{%
 \ifx #1\expandafter \@firstoftwo
 \else \expandafter \@secondoftwo
 \fi
}%
\providecommand \natexlab [1]{#1}%
\providecommand \enquote  [1]{``#1''}%
\providecommand \bibnamefont  [1]{#1}%
\providecommand \bibfnamefont [1]{#1}%
\providecommand \citenamefont [1]{#1}%
\providecommand \href@noop [0]{\@secondoftwo}%
\providecommand \href [0]{\begingroup \@sanitize@url \@href}%
\providecommand \@href[1]{\@@startlink{#1}\@@href}%
\providecommand \@@href[1]{\endgroup#1\@@endlink}%
\providecommand \@sanitize@url [0]{\catcode `\\12\catcode `\$12\catcode
  `\&12\catcode `\#12\catcode `\^12\catcode `\_12\catcode `\%12\relax}%
\providecommand \@@startlink[1]{}%
\providecommand \@@endlink[0]{}%
\providecommand \url  [0]{\begingroup\@sanitize@url \@url }%
\providecommand \@url [1]{\endgroup\@href {#1}{\urlprefix }}%
\providecommand \urlprefix  [0]{URL }%
\providecommand \Eprint [0]{\href }%
\providecommand \doibase [0]{http://dx.doi.org/}%
\providecommand \selectlanguage [0]{\@gobble}%
\providecommand \bibinfo  [0]{\@secondoftwo}%
\providecommand \bibfield  [0]{\@secondoftwo}%
\providecommand \translation [1]{[#1]}%
\providecommand \BibitemOpen [0]{}%
\providecommand \bibitemStop [0]{}%
\providecommand \bibitemNoStop [0]{.\EOS\space}%
\providecommand \EOS [0]{\spacefactor3000\relax}%
\providecommand \BibitemShut  [1]{\csname bibitem#1\endcsname}%
\let\auto@bib@innerbib\@empty
\bibitem [{\citenamefont {Myers}(2020)}]{myers2020surfactant}%
  \BibitemOpen
  \bibfield  {author} {\bibinfo {author} {\bibfnamefont {D.}~\bibnamefont
  {Myers}},\ }\href@noop {} {\emph {\bibinfo {title} {Surfactant science and
  technology}}}\ (\bibinfo  {publisher} {John Wiley \& Sons},\ \bibinfo {year}
  {2020})\BibitemShut {NoStop}%
\bibitem [{\citenamefont {Porter}(2013)}]{porter2013handbook}%
  \BibitemOpen
  \bibfield  {author} {\bibinfo {author} {\bibfnamefont {M.~R.}\ \bibnamefont
  {Porter}},\ }\href@noop {} {\emph {\bibinfo {title} {Handbook of
  surfactants}}}\ (\bibinfo  {publisher} {Springer},\ \bibinfo {year}
  {2013})\BibitemShut {NoStop}%
\bibitem [{\citenamefont {Moroi}(1992)}]{moroi1992micelles}%
  \BibitemOpen
  \bibfield  {author} {\bibinfo {author} {\bibfnamefont {Y.}~\bibnamefont
  {Moroi}},\ }\href@noop {} {\emph {\bibinfo {title} {Micelles: theoretical and
  applied aspects}}}\ (\bibinfo  {publisher} {Springer Science \& Business
  Media},\ \bibinfo {year} {1992})\BibitemShut {NoStop}%
\bibitem [{\citenamefont {Dreiss}(2007)}]{dreiss2007wormlike}%
  \BibitemOpen
  \bibfield  {author} {\bibinfo {author} {\bibfnamefont {C.~A.}\ \bibnamefont
  {Dreiss}},\ }\bibfield  {title} {\enquote {\bibinfo {title} {Wormlike
  micelles: where do we stand? recent developments, linear rheology and
  scattering techniques},}\ }\href@noop {} {\bibfield  {journal} {\bibinfo
  {journal} {Soft Matter}\ }\textbf {\bibinfo {volume} {3}},\ \bibinfo {pages}
  {956--970} (\bibinfo {year} {2007})}\BibitemShut {NoStop}%
\bibitem [{\citenamefont {Dreiss}\ and\ \citenamefont
  {Feng}(2017)}]{dreiss2017wormlike}%
  \BibitemOpen
  \bibfield  {author} {\bibinfo {author} {\bibfnamefont {C.~A.}\ \bibnamefont
  {Dreiss}}\ and\ \bibinfo {author} {\bibfnamefont {Y.}~\bibnamefont {Feng}},\
  }\href@noop {} {\emph {\bibinfo {title} {Wormlike {M}icelles: {A}dvances in
  {S}ystems, {C}haracterisation and {A}pplications}}},\ Vol.~\bibinfo {volume}
  {6}\ (\bibinfo  {publisher} {Royal Society of Chemistry},\ \bibinfo {year}
  {2017})\BibitemShut {NoStop}%
\bibitem [{\citenamefont {Anderson}, \citenamefont {Pearson},\ and\
  \citenamefont {Boek}(2006)}]{anderson2006rheology}%
  \BibitemOpen
  \bibfield  {author} {\bibinfo {author} {\bibfnamefont {V.}~\bibnamefont
  {Anderson}}, \bibinfo {author} {\bibfnamefont {J.}~\bibnamefont {Pearson}}, \
  and\ \bibinfo {author} {\bibfnamefont {E.}~\bibnamefont {Boek}},\ }\bibfield
  {title} {\enquote {\bibinfo {title} {The rheology of worm-like micellar
  fluids},}\ }\href@noop {} {\bibfield  {journal} {\bibinfo  {journal}
  {Rheology Reviews}\ }\textbf {\bibinfo {volume} {2006}},\ \bibinfo {pages}
  {217--253} (\bibinfo {year} {2006})}\BibitemShut {NoStop}%
\bibitem [{\citenamefont {Rothstein}(2003)}]{rothstein2003transient}%
  \BibitemOpen
  \bibfield  {author} {\bibinfo {author} {\bibfnamefont {J.~P.}\ \bibnamefont
  {Rothstein}},\ }\bibfield  {title} {\enquote {\bibinfo {title} {Transient
  extensional rheology of wormlike micelle solutions},}\ }\href@noop {}
  {\bibfield  {journal} {\bibinfo  {journal} {Journal of Rheology}\ }\textbf
  {\bibinfo {volume} {47}},\ \bibinfo {pages} {1227--1247} (\bibinfo {year}
  {2003})}\BibitemShut {NoStop}%
\bibitem [{\citenamefont {Rothstein}(2008)}]{rothstein2008strong}%
  \BibitemOpen
  \bibfield  {author} {\bibinfo {author} {\bibfnamefont {J.~P.}\ \bibnamefont
  {Rothstein}},\ }\bibfield  {title} {\enquote {\bibinfo {title} {Strong flows
  of viscoelastic wormlike micelle solutions},}\ }\href@noop {} {\bibfield
  {journal} {\bibinfo  {journal} {Rheology Reviews}\ }\textbf {\bibinfo
  {volume} {2008}},\ \bibinfo {pages} {1--46} (\bibinfo {year}
  {2008})}\BibitemShut {NoStop}%
\bibitem [{\citenamefont {Berret}(1997)}]{berret1997transient}%
  \BibitemOpen
  \bibfield  {author} {\bibinfo {author} {\bibfnamefont {J.-F.}\ \bibnamefont
  {Berret}},\ }\bibfield  {title} {\enquote {\bibinfo {title} {Transient
  rheology of wormlike micelles},}\ }\href@noop {} {\bibfield  {journal}
  {\bibinfo  {journal} {Langmuir}\ }\textbf {\bibinfo {volume} {13}},\ \bibinfo
  {pages} {2227--2234} (\bibinfo {year} {1997})}\BibitemShut {NoStop}%
\bibitem [{\citenamefont {Walker}(2001)}]{walker2001rheology}%
  \BibitemOpen
  \bibfield  {author} {\bibinfo {author} {\bibfnamefont {L.~M.}\ \bibnamefont
  {Walker}},\ }\bibfield  {title} {\enquote {\bibinfo {title} {Rheology and
  structure of worm-like micelles},}\ }\href@noop {} {\bibfield  {journal}
  {\bibinfo  {journal} {Current Opinion in Colloid and Interface Science}\
  }\textbf {\bibinfo {volume} {6}},\ \bibinfo {pages} {451--456} (\bibinfo
  {year} {2001})}\BibitemShut {NoStop}%
\bibitem [{\citenamefont {Yang}(2002)}]{yang2002viscoelastic}%
  \BibitemOpen
  \bibfield  {author} {\bibinfo {author} {\bibfnamefont {J.}~\bibnamefont
  {Yang}},\ }\bibfield  {title} {\enquote {\bibinfo {title} {Viscoelastic
  wormlike micelles and their applications},}\ }\href@noop {} {\bibfield
  {journal} {\bibinfo  {journal} {Current Opinion in Colloid and Interface
  Science}\ }\textbf {\bibinfo {volume} {7}},\ \bibinfo {pages} {276--281}
  (\bibinfo {year} {2002})}\BibitemShut {NoStop}%
\bibitem [{\citenamefont {Wever}, \citenamefont {Picchioni},\ and\
  \citenamefont {Broekhuis}(2011)}]{wever2011polymers}%
  \BibitemOpen
  \bibfield  {author} {\bibinfo {author} {\bibfnamefont {D.}~\bibnamefont
  {Wever}}, \bibinfo {author} {\bibfnamefont {F.}~\bibnamefont {Picchioni}}, \
  and\ \bibinfo {author} {\bibfnamefont {A.}~\bibnamefont {Broekhuis}},\
  }\bibfield  {title} {\enquote {\bibinfo {title} {Polymers for enhanced oil
  recovery: a paradigm for structure--property relationship in aqueous
  solution},}\ }\href@noop {} {\bibfield  {journal} {\bibinfo  {journal}
  {Progress in Polymer Science}\ }\textbf {\bibinfo {volume} {36}},\ \bibinfo
  {pages} {1558--1628} (\bibinfo {year} {2011})}\BibitemShut {NoStop}%
\bibitem [{\citenamefont {Raffa}, \citenamefont {Broekhuis},\ and\
  \citenamefont {Picchioni}(2016)}]{raffa2016polymeric}%
  \BibitemOpen
  \bibfield  {author} {\bibinfo {author} {\bibfnamefont {P.}~\bibnamefont
  {Raffa}}, \bibinfo {author} {\bibfnamefont {A.~A.}\ \bibnamefont
  {Broekhuis}}, \ and\ \bibinfo {author} {\bibfnamefont {F.}~\bibnamefont
  {Picchioni}},\ }\bibfield  {title} {\enquote {\bibinfo {title} {Polymeric
  surfactants for enhanced oil recovery: A review},}\ }\href@noop {} {\bibfield
   {journal} {\bibinfo  {journal} {Journal of Petroleum Science and
  Engineering}\ }\textbf {\bibinfo {volume} {145}},\ \bibinfo {pages}
  {723--733} (\bibinfo {year} {2016})}\BibitemShut {NoStop}%
\bibitem [{\citenamefont {Mandal}(2015)}]{mandal2015chemical}%
  \BibitemOpen
  \bibfield  {author} {\bibinfo {author} {\bibfnamefont {A.}~\bibnamefont
  {Mandal}},\ }\bibfield  {title} {\enquote {\bibinfo {title} {Chemical flood
  enhanced oil recovery: a review},}\ }\href@noop {} {\bibfield  {journal}
  {\bibinfo  {journal} {International Journal of Oil, Gas and Coal Technology}\
  }\textbf {\bibinfo {volume} {9}},\ \bibinfo {pages} {241--264} (\bibinfo
  {year} {2015})}\BibitemShut {NoStop}%
\bibitem [{\citenamefont {Mosler}\ and\ \citenamefont
  {Hatton}(1996)}]{mosler1996surfactants}%
  \BibitemOpen
  \bibfield  {author} {\bibinfo {author} {\bibfnamefont {R.}~\bibnamefont
  {Mosler}}\ and\ \bibinfo {author} {\bibfnamefont {T.~A.}\ \bibnamefont
  {Hatton}},\ }\bibfield  {title} {\enquote {\bibinfo {title} {Surfactants and
  polymers for environmental remediation and control},}\ }\href@noop {}
  {\bibfield  {journal} {\bibinfo  {journal} {Current Opinion in Colloid \&
  Interface Science}\ }\textbf {\bibinfo {volume} {1}},\ \bibinfo {pages}
  {540--547} (\bibinfo {year} {1996})}\BibitemShut {NoStop}%
\bibitem [{\citenamefont {Dwarakanath}\ \emph {et~al.}(1999)\citenamefont
  {Dwarakanath}, \citenamefont {Kostarelos}, \citenamefont {Pope},
  \citenamefont {Shotts},\ and\ \citenamefont {Wade}}]{dwarakanath1999anionic}%
  \BibitemOpen
  \bibfield  {author} {\bibinfo {author} {\bibfnamefont {V.}~\bibnamefont
  {Dwarakanath}}, \bibinfo {author} {\bibfnamefont {K.}~\bibnamefont
  {Kostarelos}}, \bibinfo {author} {\bibfnamefont {G.~A.}\ \bibnamefont
  {Pope}}, \bibinfo {author} {\bibfnamefont {D.}~\bibnamefont {Shotts}}, \ and\
  \bibinfo {author} {\bibfnamefont {W.~H.}\ \bibnamefont {Wade}},\ }\bibfield
  {title} {\enquote {\bibinfo {title} {Anionic surfactant remediation of soil
  columns contaminated by nonaqueous phase liquids},}\ }\href@noop {}
  {\bibfield  {journal} {\bibinfo  {journal} {Journal of Contaminant
  Hydrology}\ }\textbf {\bibinfo {volume} {38}},\ \bibinfo {pages} {465--488}
  (\bibinfo {year} {1999})}\BibitemShut {NoStop}%
\bibitem [{\citenamefont {Darcy}(1856)}]{darcy1856fontaines}%
  \BibitemOpen
  \bibfield  {author} {\bibinfo {author} {\bibfnamefont {H.}~\bibnamefont
  {Darcy}},\ }\href@noop {} {\emph {\bibinfo {title} {Les fontaines publiques
  de la ville de Dijon: exposition et application des principes a suivre et des
  formules a}}}\ (\bibinfo  {publisher} {Victor Dalmont},\ \bibinfo {year}
  {1856})\BibitemShut {NoStop}%
\bibitem [{\citenamefont {Carman}(1997)}]{carman1997fluid}%
  \BibitemOpen
  \bibfield  {author} {\bibinfo {author} {\bibfnamefont {P.~C.}\ \bibnamefont
  {Carman}},\ }\bibfield  {title} {\enquote {\bibinfo {title} {Fluid flow
  through granular beds},}\ }\href@noop {} {\bibfield  {journal} {\bibinfo
  {journal} {Chemical Engineering Research and Design}\ }\textbf {\bibinfo
  {volume} {75}},\ \bibinfo {pages} {S32--S48} (\bibinfo {year}
  {1997})}\BibitemShut {NoStop}%
\bibitem [{\citenamefont {Ergun}\ and\ \citenamefont
  {Orning}(1949)}]{ergun1949fluid}%
  \BibitemOpen
  \bibfield  {author} {\bibinfo {author} {\bibfnamefont {S.}~\bibnamefont
  {Ergun}}\ and\ \bibinfo {author} {\bibfnamefont {A.~A.}\ \bibnamefont
  {Orning}},\ }\bibfield  {title} {\enquote {\bibinfo {title} {Fluid flow
  through randomly packed columns and fluidized beds},}\ }\href@noop {}
  {\bibfield  {journal} {\bibinfo  {journal} {Industrial \& Engineering
  Chemistry Research}\ }\textbf {\bibinfo {volume} {41}},\ \bibinfo {pages}
  {1179--1184} (\bibinfo {year} {1949})}\BibitemShut {NoStop}%
\bibitem [{\citenamefont {Sochi}(2010)}]{sochi2010non}%
  \BibitemOpen
  \bibfield  {author} {\bibinfo {author} {\bibfnamefont {T.}~\bibnamefont
  {Sochi}},\ }\bibfield  {title} {\enquote {\bibinfo {title} {Non-newtonian
  flow in porous media},}\ }\href@noop {} {\bibfield  {journal} {\bibinfo
  {journal} {Polymer}\ }\textbf {\bibinfo {volume} {51}},\ \bibinfo {pages}
  {5007--5023} (\bibinfo {year} {2010})}\BibitemShut {NoStop}%
\bibitem [{\citenamefont {Marshall}\ and\ \citenamefont
  {Metzner}(1967)}]{marshall1967flow}%
  \BibitemOpen
  \bibfield  {author} {\bibinfo {author} {\bibfnamefont {R.}~\bibnamefont
  {Marshall}}\ and\ \bibinfo {author} {\bibfnamefont {A.}~\bibnamefont
  {Metzner}},\ }\bibfield  {title} {\enquote {\bibinfo {title} {Flow of
  viscoelastic fluids through porous media},}\ }\href@noop {} {\bibfield
  {journal} {\bibinfo  {journal} {Industrial \& Engineering Chemistry
  Fundamentals}\ }\textbf {\bibinfo {volume} {6}},\ \bibinfo {pages} {393--400}
  (\bibinfo {year} {1967})}\BibitemShut {NoStop}%
\bibitem [{\citenamefont {Deiber}\ and\ \citenamefont
  {Schowalter}(1981)}]{deiber1981modeling}%
  \BibitemOpen
  \bibfield  {author} {\bibinfo {author} {\bibfnamefont {J.}~\bibnamefont
  {Deiber}}\ and\ \bibinfo {author} {\bibfnamefont {W.}~\bibnamefont
  {Schowalter}},\ }\bibfield  {title} {\enquote {\bibinfo {title} {Modeling the
  flow of viscoelastic fluids through porous media},}\ }\href@noop {}
  {\bibfield  {journal} {\bibinfo  {journal} {AIChE Journal}\ }\textbf
  {\bibinfo {volume} {27}},\ \bibinfo {pages} {912--920} (\bibinfo {year}
  {1981})}\BibitemShut {NoStop}%
\bibitem [{\citenamefont {Slattery}(1967)}]{slattery1967flow}%
  \BibitemOpen
  \bibfield  {author} {\bibinfo {author} {\bibfnamefont {J.~C.}\ \bibnamefont
  {Slattery}},\ }\bibfield  {title} {\enquote {\bibinfo {title} {Flow of
  viscoelastic fluids through porous media},}\ }\href@noop {} {\bibfield
  {journal} {\bibinfo  {journal} {AIChE Journal}\ }\textbf {\bibinfo {volume}
  {13}},\ \bibinfo {pages} {1066--1071} (\bibinfo {year} {1967})}\BibitemShut
  {NoStop}%
\bibitem [{\citenamefont {Sobti}\ and\ \citenamefont
  {Wanchoo}(2014)}]{sobti2014creeping}%
  \BibitemOpen
  \bibfield  {author} {\bibinfo {author} {\bibfnamefont {A.}~\bibnamefont
  {Sobti}}\ and\ \bibinfo {author} {\bibfnamefont {R.~K.}\ \bibnamefont
  {Wanchoo}},\ }\bibfield  {title} {\enquote {\bibinfo {title} {Creeping flow
  of viscoelastic fluid through a packed bed},}\ }\href@noop {} {\bibfield
  {journal} {\bibinfo  {journal} {Industrial \& Engineering Chemistry
  Research}\ }\textbf {\bibinfo {volume} {53}},\ \bibinfo {pages}
  {14508--14518} (\bibinfo {year} {2014})}\BibitemShut {NoStop}%
\bibitem [{\citenamefont {Tiu}\ \emph {et~al.}(1997)\citenamefont {Tiu},
  \citenamefont {Zhou}, \citenamefont {Nicolae}, \citenamefont {Fang},\ and\
  \citenamefont {Chhabra}}]{tiu1997flow}%
  \BibitemOpen
  \bibfield  {author} {\bibinfo {author} {\bibfnamefont {C.}~\bibnamefont
  {Tiu}}, \bibinfo {author} {\bibfnamefont {J.~Z.}\ \bibnamefont {Zhou}},
  \bibinfo {author} {\bibfnamefont {G.}~\bibnamefont {Nicolae}}, \bibinfo
  {author} {\bibfnamefont {T.}~\bibnamefont {Fang}}, \ and\ \bibinfo {author}
  {\bibfnamefont {R.~P.}\ \bibnamefont {Chhabra}},\ }\bibfield  {title}
  {\enquote {\bibinfo {title} {Flow of viscoelastic polymer solutions in mixed
  beds of particles},}\ }\href@noop {} {\bibfield  {journal} {\bibinfo
  {journal} {The Canadian Journal of Chemical Engineering}\ }\textbf {\bibinfo
  {volume} {75}},\ \bibinfo {pages} {843--850} (\bibinfo {year}
  {1997})}\BibitemShut {NoStop}%
\bibitem [{\citenamefont {Browne}, \citenamefont {Shih},\ and\ \citenamefont
  {Datta}(2020{\natexlab{a}})}]{browne2020pore}%
  \BibitemOpen
  \bibfield  {author} {\bibinfo {author} {\bibfnamefont {C.~A.}\ \bibnamefont
  {Browne}}, \bibinfo {author} {\bibfnamefont {A.}~\bibnamefont {Shih}}, \ and\
  \bibinfo {author} {\bibfnamefont {S.~S.}\ \bibnamefont {Datta}},\ }\bibfield
  {title} {\enquote {\bibinfo {title} {Pore-scale flow characterization of
  polymer solutions in microfluidic porous media},}\ }\href@noop {} {\bibfield
  {journal} {\bibinfo  {journal} {Small}\ }\textbf {\bibinfo {volume} {16}},\
  \bibinfo {pages} {1903944} (\bibinfo {year}
  {2020}{\natexlab{a}})}\BibitemShut {NoStop}%
\bibitem [{\citenamefont {De}\ \emph {et~al.}(2017{\natexlab{a}})\citenamefont
  {De}, \citenamefont {Van Der~Schaaf}, \citenamefont {Deen}, \citenamefont
  {Kuipers}, \citenamefont {Peters},\ and\ \citenamefont
  {Padding}}]{de2017lane}%
  \BibitemOpen
  \bibfield  {author} {\bibinfo {author} {\bibfnamefont {S.}~\bibnamefont
  {De}}, \bibinfo {author} {\bibfnamefont {J.}~\bibnamefont {Van Der~Schaaf}},
  \bibinfo {author} {\bibfnamefont {N.~G.}\ \bibnamefont {Deen}}, \bibinfo
  {author} {\bibfnamefont {J.}~\bibnamefont {Kuipers}}, \bibinfo {author}
  {\bibfnamefont {E.}~\bibnamefont {Peters}}, \ and\ \bibinfo {author}
  {\bibfnamefont {J.}~\bibnamefont {Padding}},\ }\bibfield  {title} {\enquote
  {\bibinfo {title} {Lane change in flows through pillared microchannels},}\
  }\href@noop {} {\bibfield  {journal} {\bibinfo  {journal} {Physics of
  Fluids}\ }\textbf {\bibinfo {volume} {29}},\ \bibinfo {pages} {113102}
  (\bibinfo {year} {2017}{\natexlab{a}})}\BibitemShut {NoStop}%
\bibitem [{\citenamefont {De}\ \emph {et~al.}(2018)\citenamefont {De},
  \citenamefont {Koesen}, \citenamefont {Maitri}, \citenamefont {Golombok},
  \citenamefont {Padding},\ and\ \citenamefont {van Santvoort}}]{de2018flow}%
  \BibitemOpen
  \bibfield  {author} {\bibinfo {author} {\bibfnamefont {S.}~\bibnamefont
  {De}}, \bibinfo {author} {\bibfnamefont {S.}~\bibnamefont {Koesen}}, \bibinfo
  {author} {\bibfnamefont {R.}~\bibnamefont {Maitri}}, \bibinfo {author}
  {\bibfnamefont {M.}~\bibnamefont {Golombok}}, \bibinfo {author}
  {\bibfnamefont {J.}~\bibnamefont {Padding}}, \ and\ \bibinfo {author}
  {\bibfnamefont {J.}~\bibnamefont {van Santvoort}},\ }\bibfield  {title}
  {\enquote {\bibinfo {title} {Flow of viscoelastic surfactants through porous
  media},}\ }\href@noop {} {\bibfield  {journal} {\bibinfo  {journal} {AIChE
  Journal}\ }\textbf {\bibinfo {volume} {64}},\ \bibinfo {pages} {773--781}
  (\bibinfo {year} {2018})}\BibitemShut {NoStop}%
\bibitem [{\citenamefont {Walkama}, \citenamefont {Waisbord},\ and\
  \citenamefont {Guasto}(2020)}]{walkama2020disorder}%
  \BibitemOpen
  \bibfield  {author} {\bibinfo {author} {\bibfnamefont {D.~M.}\ \bibnamefont
  {Walkama}}, \bibinfo {author} {\bibfnamefont {N.}~\bibnamefont {Waisbord}}, \
  and\ \bibinfo {author} {\bibfnamefont {J.~S.}\ \bibnamefont {Guasto}},\
  }\bibfield  {title} {\enquote {\bibinfo {title} {Disorder suppresses chaos in
  viscoelastic flows},}\ }\href@noop {} {\bibfield  {journal} {\bibinfo
  {journal} {Physical Review Letters}\ }\textbf {\bibinfo {volume} {124}},\
  \bibinfo {pages} {164501} (\bibinfo {year} {2020})}\BibitemShut {NoStop}%
\bibitem [{\citenamefont {Haward}, \citenamefont {Hopkins},\ and\ \citenamefont
  {Shen}(2021)}]{haward2021stagnation}%
  \BibitemOpen
  \bibfield  {author} {\bibinfo {author} {\bibfnamefont {S.~J.}\ \bibnamefont
  {Haward}}, \bibinfo {author} {\bibfnamefont {C.~C.}\ \bibnamefont {Hopkins}},
  \ and\ \bibinfo {author} {\bibfnamefont {A.~Q.}\ \bibnamefont {Shen}},\
  }\bibfield  {title} {\enquote {\bibinfo {title} {Stagnation points control
  chaotic fluctuations in viscoelastic porous media flow},}\ }\href@noop {}
  {\bibfield  {journal} {\bibinfo  {journal} {arXiv preprint arXiv:2105.11063}\
  } (\bibinfo {year} {2021})}\BibitemShut {NoStop}%
\bibitem [{\citenamefont {Galindo-Rosales}\ \emph {et~al.}(2012)\citenamefont
  {Galindo-Rosales}, \citenamefont {Campo-Deano}, \citenamefont {Pinho},
  \citenamefont {Van~Bokhorst}, \citenamefont {Hamersma}, \citenamefont
  {Oliveira},\ and\ \citenamefont {Alves}}]{galindo2012microfluidic}%
  \BibitemOpen
  \bibfield  {author} {\bibinfo {author} {\bibfnamefont {F.~J.}\ \bibnamefont
  {Galindo-Rosales}}, \bibinfo {author} {\bibfnamefont {L.}~\bibnamefont
  {Campo-Deano}}, \bibinfo {author} {\bibfnamefont {F.~T.}\ \bibnamefont
  {Pinho}}, \bibinfo {author} {\bibfnamefont {E.}~\bibnamefont {Van~Bokhorst}},
  \bibinfo {author} {\bibfnamefont {P.}~\bibnamefont {Hamersma}}, \bibinfo
  {author} {\bibfnamefont {M.~S.}\ \bibnamefont {Oliveira}}, \ and\ \bibinfo
  {author} {\bibfnamefont {M.~A.}\ \bibnamefont {Alves}},\ }\bibfield  {title}
  {\enquote {\bibinfo {title} {Microfluidic systems for the analysis of
  viscoelastic fluid flow phenomena in porous media},}\ }\href@noop {}
  {\bibfield  {journal} {\bibinfo  {journal} {Microfluidics and Nanofluidics}\
  }\textbf {\bibinfo {volume} {12}},\ \bibinfo {pages} {485--498} (\bibinfo
  {year} {2012})}\BibitemShut {NoStop}%
\bibitem [{\citenamefont {Browne}, \citenamefont {Shih},\ and\ \citenamefont
  {Datta}(2020{\natexlab{b}})}]{browne2020bistability}%
  \BibitemOpen
  \bibfield  {author} {\bibinfo {author} {\bibfnamefont {C.~A.}\ \bibnamefont
  {Browne}}, \bibinfo {author} {\bibfnamefont {A.}~\bibnamefont {Shih}}, \ and\
  \bibinfo {author} {\bibfnamefont {S.~S.}\ \bibnamefont {Datta}},\ }\bibfield
  {title} {\enquote {\bibinfo {title} {Bistability in the unstable flow of
  polymer solutions through pore constriction arrays},}\ }\href@noop {}
  {\bibfield  {journal} {\bibinfo  {journal} {Journal of Fluid Mechanics}\
  }\textbf {\bibinfo {volume} {890}} (\bibinfo {year}
  {2020}{\natexlab{b}})}\BibitemShut {NoStop}%
\bibitem [{\citenamefont {Ekanem}\ \emph {et~al.}(2020)\citenamefont {Ekanem},
  \citenamefont {Berg}, \citenamefont {De}, \citenamefont {Fadili},
  \citenamefont {Bultreys}, \citenamefont {R{\"u}cker}, \citenamefont
  {Southwick}, \citenamefont {Crawshaw},\ and\ \citenamefont
  {Luckham}}]{ekanem2020signature}%
  \BibitemOpen
  \bibfield  {author} {\bibinfo {author} {\bibfnamefont {E.~M.}\ \bibnamefont
  {Ekanem}}, \bibinfo {author} {\bibfnamefont {S.}~\bibnamefont {Berg}},
  \bibinfo {author} {\bibfnamefont {S.}~\bibnamefont {De}}, \bibinfo {author}
  {\bibfnamefont {A.}~\bibnamefont {Fadili}}, \bibinfo {author} {\bibfnamefont
  {T.}~\bibnamefont {Bultreys}}, \bibinfo {author} {\bibfnamefont
  {M.}~\bibnamefont {R{\"u}cker}}, \bibinfo {author} {\bibfnamefont
  {J.}~\bibnamefont {Southwick}}, \bibinfo {author} {\bibfnamefont
  {J.}~\bibnamefont {Crawshaw}}, \ and\ \bibinfo {author} {\bibfnamefont
  {P.~F.}\ \bibnamefont {Luckham}},\ }\bibfield  {title} {\enquote {\bibinfo
  {title} {Signature of elastic turbulence of viscoelastic fluid flow in a
  single pore throat},}\ }\href@noop {} {\bibfield  {journal} {\bibinfo
  {journal} {Physical Review E}\ }\textbf {\bibinfo {volume} {101}},\ \bibinfo
  {pages} {042605} (\bibinfo {year} {2020})}\BibitemShut {NoStop}%
\bibitem [{\citenamefont {Kumar}\ \emph {et~al.}(2021)\citenamefont {Kumar},
  \citenamefont {Aramideh}, \citenamefont {Browne}, \citenamefont {Datta},\
  and\ \citenamefont {Ardekani}}]{kumar2021numerical}%
  \BibitemOpen
  \bibfield  {author} {\bibinfo {author} {\bibfnamefont {M.}~\bibnamefont
  {Kumar}}, \bibinfo {author} {\bibfnamefont {S.}~\bibnamefont {Aramideh}},
  \bibinfo {author} {\bibfnamefont {C.~A.}\ \bibnamefont {Browne}}, \bibinfo
  {author} {\bibfnamefont {S.~S.}\ \bibnamefont {Datta}}, \ and\ \bibinfo
  {author} {\bibfnamefont {A.~M.}\ \bibnamefont {Ardekani}},\ }\bibfield
  {title} {\enquote {\bibinfo {title} {Numerical investigation of
  multistability in the unstable flow of a polymer solution through porous
  media},}\ }\href@noop {} {\bibfield  {journal} {\bibinfo  {journal} {Physical
  Review Fluids}\ }\textbf {\bibinfo {volume} {6}},\ \bibinfo {pages} {033304}
  (\bibinfo {year} {2021})}\BibitemShut {NoStop}%
\bibitem [{\citenamefont {De}\ \emph {et~al.}(2017{\natexlab{b}})\citenamefont
  {De}, \citenamefont {Kuipers}, \citenamefont {Peters},\ and\ \citenamefont
  {Padding}}]{de2017viscoelastic}%
  \BibitemOpen
  \bibfield  {author} {\bibinfo {author} {\bibfnamefont {S.}~\bibnamefont
  {De}}, \bibinfo {author} {\bibfnamefont {J.}~\bibnamefont {Kuipers}},
  \bibinfo {author} {\bibfnamefont {E.}~\bibnamefont {Peters}}, \ and\ \bibinfo
  {author} {\bibfnamefont {J.}~\bibnamefont {Padding}},\ }\bibfield  {title}
  {\enquote {\bibinfo {title} {Viscoelastic flow simulations in random porous
  media},}\ }\href@noop {} {\bibfield  {journal} {\bibinfo  {journal} {Journal
  of Non-Newtonian Fluid Mechanics}\ }\textbf {\bibinfo {volume} {248}},\
  \bibinfo {pages} {50--61} (\bibinfo {year} {2017}{\natexlab{b}})}\BibitemShut
  {NoStop}%
\bibitem [{\citenamefont {Kalb}, \citenamefont {Villasmil-Urdaneta},\ and\
  \citenamefont {Cromer}(2018)}]{kalb2018elastic}%
  \BibitemOpen
  \bibfield  {author} {\bibinfo {author} {\bibfnamefont {A.}~\bibnamefont
  {Kalb}}, \bibinfo {author} {\bibfnamefont {L.~A.}\ \bibnamefont
  {Villasmil-Urdaneta}}, \ and\ \bibinfo {author} {\bibfnamefont
  {M.}~\bibnamefont {Cromer}},\ }\bibfield  {title} {\enquote {\bibinfo {title}
  {Elastic instability and secondary flow in cross-slot flow of wormlike
  micellar solutions},}\ }\href@noop {} {\bibfield  {journal} {\bibinfo
  {journal} {Journal of Non-Newtonian Fluid Mechanics}\ }\textbf {\bibinfo
  {volume} {262}},\ \bibinfo {pages} {79--91} (\bibinfo {year}
  {2018})}\BibitemShut {NoStop}%
\bibitem [{\citenamefont {Kalb}, \citenamefont {U},\ and\ \citenamefont
  {Cromer}(2017)}]{kalb2017role}%
  \BibitemOpen
  \bibfield  {author} {\bibinfo {author} {\bibfnamefont {A.}~\bibnamefont
  {Kalb}}, \bibinfo {author} {\bibfnamefont {L.~A.~V.}\ \bibnamefont {U}}, \
  and\ \bibinfo {author} {\bibfnamefont {M.}~\bibnamefont {Cromer}},\
  }\bibfield  {title} {\enquote {\bibinfo {title} {Role of chain scission in
  cross-slot flow of wormlike micellar solutions},}\ }\href@noop {} {\bibfield
  {journal} {\bibinfo  {journal} {Physical Review Fluids}\ }\textbf {\bibinfo
  {volume} {2}},\ \bibinfo {pages} {071301} (\bibinfo {year}
  {2017})}\BibitemShut {NoStop}%
\bibitem [{\citenamefont {Sasmal}(2021)}]{sasmal2021unsteady}%
  \BibitemOpen
  \bibfield  {author} {\bibinfo {author} {\bibfnamefont {C.}~\bibnamefont
  {Sasmal}},\ }\bibfield  {title} {\enquote {\bibinfo {title} {Unsteady motion
  past a sphere translating steadily in wormlike micellar solutions: A
  numerical analysis},}\ }\href@noop {} {\bibfield  {journal} {\bibinfo
  {journal} {Journal of Fluid Mechanics}\ }\textbf {\bibinfo {volume} {912}}
  (\bibinfo {year} {2021})}\BibitemShut {NoStop}%
\bibitem [{\citenamefont {Sasmal}(2020)}]{sasmal2020flow}%
  \BibitemOpen
  \bibfield  {author} {\bibinfo {author} {\bibfnamefont {C.}~\bibnamefont
  {Sasmal}},\ }\bibfield  {title} {\enquote {\bibinfo {title} {Flow of wormlike
  micellar solutions through a long micropore with step expansion and
  contraction},}\ }\href@noop {} {\bibfield  {journal} {\bibinfo  {journal}
  {Physics of Fluids}\ }\textbf {\bibinfo {volume} {32}},\ \bibinfo {pages}
  {013103} (\bibinfo {year} {2020})}\BibitemShut {NoStop}%
\bibitem [{\citenamefont {Mohammadigoushki}\ \emph {et~al.}(2019)\citenamefont
  {Mohammadigoushki}, \citenamefont {Dalili}, \citenamefont {Zhou},\ and\
  \citenamefont {Cook}}]{mohammadigoushki2019transient}%
  \BibitemOpen
  \bibfield  {author} {\bibinfo {author} {\bibfnamefont {H.}~\bibnamefont
  {Mohammadigoushki}}, \bibinfo {author} {\bibfnamefont {A.}~\bibnamefont
  {Dalili}}, \bibinfo {author} {\bibfnamefont {L.}~\bibnamefont {Zhou}}, \ and\
  \bibinfo {author} {\bibfnamefont {P.}~\bibnamefont {Cook}},\ }\bibfield
  {title} {\enquote {\bibinfo {title} {Transient evolution of flow profiles in
  a shear banding wormlike micellar solution: Experimental results and a
  comparison with the {VCM} model},}\ }\href@noop {} {\bibfield  {journal}
  {\bibinfo  {journal} {Soft matter}\ }\textbf {\bibinfo {volume} {15}},\
  \bibinfo {pages} {5483--5494} (\bibinfo {year} {2019})}\BibitemShut {NoStop}%
\bibitem [{\citenamefont {Vasquez}, \citenamefont {McKinley},\ and\
  \citenamefont {Cook}(2007)}]{vasquez2007network}%
  \BibitemOpen
  \bibfield  {author} {\bibinfo {author} {\bibfnamefont {P.~A.}\ \bibnamefont
  {Vasquez}}, \bibinfo {author} {\bibfnamefont {G.~H.}\ \bibnamefont
  {McKinley}}, \ and\ \bibinfo {author} {\bibfnamefont {L.~P.}\ \bibnamefont
  {Cook}},\ }\bibfield  {title} {\enquote {\bibinfo {title} {A network scission
  model for wormlike micellar solutions: I. model formulation and viscometric
  flow predictions},}\ }\href@noop {} {\bibfield  {journal} {\bibinfo
  {journal} {Journal of non-newtonian fluid mechanics}\ }\textbf {\bibinfo
  {volume} {144}},\ \bibinfo {pages} {122--139} (\bibinfo {year}
  {2007})}\BibitemShut {NoStop}%
\bibitem [{\citenamefont {Cates}(1987)}]{cates1987reptation}%
  \BibitemOpen
  \bibfield  {author} {\bibinfo {author} {\bibfnamefont {M.~E.}\ \bibnamefont
  {Cates}},\ }\bibfield  {title} {\enquote {\bibinfo {title} {Reptation of
  living polymers: dynamics of entangled polymers in the presence of reversible
  chain-scission reactions},}\ }\href@noop {} {\bibfield  {journal} {\bibinfo
  {journal} {Macromolecules}\ }\textbf {\bibinfo {volume} {20}},\ \bibinfo
  {pages} {2289--2296} (\bibinfo {year} {1987})}\BibitemShut {NoStop}%
\bibitem [{\citenamefont {Pipe}\ \emph {et~al.}(2010)\citenamefont {Pipe},
  \citenamefont {Kim}, \citenamefont {Vasquez}, \citenamefont {Cook},\ and\
  \citenamefont {McKinley}}]{pipe2010wormlike}%
  \BibitemOpen
  \bibfield  {author} {\bibinfo {author} {\bibfnamefont {C.}~\bibnamefont
  {Pipe}}, \bibinfo {author} {\bibfnamefont {N.}~\bibnamefont {Kim}}, \bibinfo
  {author} {\bibfnamefont {P.}~\bibnamefont {Vasquez}}, \bibinfo {author}
  {\bibfnamefont {L.}~\bibnamefont {Cook}}, \ and\ \bibinfo {author}
  {\bibfnamefont {G.}~\bibnamefont {McKinley}},\ }\bibfield  {title} {\enquote
  {\bibinfo {title} {Wormlike micellar solutions: Ii. comparison between
  experimental data and scission model predictions},}\ }\href@noop {}
  {\bibfield  {journal} {\bibinfo  {journal} {Journal of Rheology}\ }\textbf
  {\bibinfo {volume} {54}},\ \bibinfo {pages} {881--913} (\bibinfo {year}
  {2010})}\BibitemShut {NoStop}%
\bibitem [{\citenamefont {Zhou}, \citenamefont {McKinley},\ and\ \citenamefont
  {Cook}(2014)}]{zhou2014wormlike}%
  \BibitemOpen
  \bibfield  {author} {\bibinfo {author} {\bibfnamefont {L.}~\bibnamefont
  {Zhou}}, \bibinfo {author} {\bibfnamefont {G.~H.}\ \bibnamefont {McKinley}},
  \ and\ \bibinfo {author} {\bibfnamefont {L.~P.}\ \bibnamefont {Cook}},\
  }\bibfield  {title} {\enquote {\bibinfo {title} {Wormlike micellar solutions:
  Iii. {VCM} model predictions in steady and transient shearing flows},}\
  }\href@noop {} {\bibfield  {journal} {\bibinfo  {journal} {Journal of
  Non-Newtonian Fluid Mechanics}\ }\textbf {\bibinfo {volume} {211}},\ \bibinfo
  {pages} {70--83} (\bibinfo {year} {2014})}\BibitemShut {NoStop}%
\bibitem [{\citenamefont {Cromer}, \citenamefont {Cook},\ and\ \citenamefont
  {McKinley}(2009)}]{cromer2009extensional}%
  \BibitemOpen
  \bibfield  {author} {\bibinfo {author} {\bibfnamefont {M.}~\bibnamefont
  {Cromer}}, \bibinfo {author} {\bibfnamefont {L.~P.}\ \bibnamefont {Cook}}, \
  and\ \bibinfo {author} {\bibfnamefont {G.~H.}\ \bibnamefont {McKinley}},\
  }\bibfield  {title} {\enquote {\bibinfo {title} {Extensional flow of wormlike
  micellar solutions},}\ }\href@noop {} {\bibfield  {journal} {\bibinfo
  {journal} {Chemical Engineering Science}\ }\textbf {\bibinfo {volume} {64}},\
  \bibinfo {pages} {4588--4596} (\bibinfo {year} {2009})}\BibitemShut {NoStop}%
\bibitem [{\citenamefont {Olmsted}, \citenamefont {Radulescu},\ and\
  \citenamefont {Lu}(2000)}]{olmsted2000johnson}%
  \BibitemOpen
  \bibfield  {author} {\bibinfo {author} {\bibfnamefont {P.}~\bibnamefont
  {Olmsted}}, \bibinfo {author} {\bibfnamefont {O.}~\bibnamefont {Radulescu}},
  \ and\ \bibinfo {author} {\bibfnamefont {C.-Y.}\ \bibnamefont {Lu}},\
  }\bibfield  {title} {\enquote {\bibinfo {title} {Johnson--segalman model with
  a diffusion term in cylindrical couette flow},}\ }\href@noop {} {\bibfield
  {journal} {\bibinfo  {journal} {Journal of Rheology}\ }\textbf {\bibinfo
  {volume} {44}},\ \bibinfo {pages} {257--275} (\bibinfo {year}
  {2000})}\BibitemShut {NoStop}%
\bibitem [{\citenamefont {Lu}, \citenamefont {Olmsted},\ and\ \citenamefont
  {Ball}(2000)}]{lu2000effects}%
  \BibitemOpen
  \bibfield  {author} {\bibinfo {author} {\bibfnamefont {C.-Y.~D.}\
  \bibnamefont {Lu}}, \bibinfo {author} {\bibfnamefont {P.~D.}\ \bibnamefont
  {Olmsted}}, \ and\ \bibinfo {author} {\bibfnamefont {R.}~\bibnamefont
  {Ball}},\ }\bibfield  {title} {\enquote {\bibinfo {title} {Effects of
  nonlocal stress on the determination of shear banding flow},}\ }\href@noop {}
  {\bibfield  {journal} {\bibinfo  {journal} {Physical Review Letters}\
  }\textbf {\bibinfo {volume} {84}},\ \bibinfo {pages} {642--645} (\bibinfo
  {year} {2000})}\BibitemShut {NoStop}%
\bibitem [{\citenamefont {Bautista}\ \emph {et~al.}(1999)\citenamefont
  {Bautista}, \citenamefont {De~Santos}, \citenamefont {Puig},\ and\
  \citenamefont {Manero}}]{bautista1999understanding}%
  \BibitemOpen
  \bibfield  {author} {\bibinfo {author} {\bibfnamefont {F.}~\bibnamefont
  {Bautista}}, \bibinfo {author} {\bibfnamefont {J.}~\bibnamefont {De~Santos}},
  \bibinfo {author} {\bibfnamefont {J.}~\bibnamefont {Puig}}, \ and\ \bibinfo
  {author} {\bibfnamefont {O.}~\bibnamefont {Manero}},\ }\bibfield  {title}
  {\enquote {\bibinfo {title} {Understanding thixotropic and antithixotropic
  behavior of viscoelastic micellar solutions and liquid crystalline
  dispersions i the model},}\ }\href@noop {} {\bibfield  {journal} {\bibinfo
  {journal} {Journal of Non-Newtonian Fluid Mechanics}\ }\textbf {\bibinfo
  {volume} {80}},\ \bibinfo {pages} {93--113} (\bibinfo {year}
  {1999})}\BibitemShut {NoStop}%
\bibitem [{\citenamefont {Khan}\ and\ \citenamefont
  {Sasmal}(2020)}]{khan2020effect}%
  \BibitemOpen
  \bibfield  {author} {\bibinfo {author} {\bibfnamefont {M.~B.}\ \bibnamefont
  {Khan}}\ and\ \bibinfo {author} {\bibfnamefont {C.}~\bibnamefont {Sasmal}},\
  }\bibfield  {title} {\enquote {\bibinfo {title} {Effect of chain scission on
  flow characteristics of wormlike micellar solutions past a confined
  microfluidic cylinder: A numerical analysis},}\ }\href@noop {} {\bibfield
  {journal} {\bibinfo  {journal} {Soft Matter}\ }\textbf {\bibinfo {volume}
  {16}},\ \bibinfo {pages} {5261--5272} (\bibinfo {year} {2020})}\BibitemShut
  {NoStop}%
\bibitem [{\citenamefont {Weller}\ \emph {et~al.}(1998)\citenamefont {Weller},
  \citenamefont {Tabor}, \citenamefont {Jasak},\ and\ \citenamefont
  {Fureby}}]{wellerOpenFOAM}%
  \BibitemOpen
  \bibfield  {author} {\bibinfo {author} {\bibfnamefont {H.~G.}\ \bibnamefont
  {Weller}}, \bibinfo {author} {\bibfnamefont {G.}~\bibnamefont {Tabor}},
  \bibinfo {author} {\bibfnamefont {H.}~\bibnamefont {Jasak}}, \ and\ \bibinfo
  {author} {\bibfnamefont {C.}~\bibnamefont {Fureby}},\ }\bibfield  {title}
  {\enquote {\bibinfo {title} {A tensorial approach to computational continuum
  mechanics using object-oriented techniques},}\ }\href@noop {} {\bibfield
  {journal} {\bibinfo  {journal} {Computers in Physics}\ }\textbf {\bibinfo
  {volume} {12}},\ \bibinfo {pages} {620--631} (\bibinfo {year}
  {1998})}\BibitemShut {NoStop}%
\bibitem [{\citenamefont {Pimenta}\ and\ \citenamefont
  {Alves}(2016)}]{rheoTool}%
  \BibitemOpen
  \bibfield  {author} {\bibinfo {author} {\bibfnamefont {F.}~\bibnamefont
  {Pimenta}}\ and\ \bibinfo {author} {\bibfnamefont {M.}~\bibnamefont
  {Alves}},\ }\href@noop {} {\enquote {\bibinfo {title} {rheotool},}\ }\bibinfo
  {howpublished} {\url{https://github.com/fppimenta/rheoTool}} (\bibinfo {year}
  {2016})\BibitemShut {NoStop}%
\bibitem [{\citenamefont {Ajiz}\ and\ \citenamefont
  {Jennings}(1984)}]{ajiz1984robust}%
  \BibitemOpen
  \bibfield  {author} {\bibinfo {author} {\bibfnamefont {M.~A.}\ \bibnamefont
  {Ajiz}}\ and\ \bibinfo {author} {\bibfnamefont {A.}~\bibnamefont
  {Jennings}},\ }\bibfield  {title} {\enquote {\bibinfo {title} {A robust
  incomplete choleski-conjugate gradient algorithm},}\ }\href@noop {}
  {\bibfield  {journal} {\bibinfo  {journal} {International Journal for
  Numerical Methods in Engineering}\ }\textbf {\bibinfo {volume} {20}},\
  \bibinfo {pages} {949--966} (\bibinfo {year} {1984})}\BibitemShut {NoStop}%
\bibitem [{\citenamefont {Lee}, \citenamefont {Zhang},\ and\ \citenamefont
  {Lu}(2003)}]{lee2003incomplete}%
  \BibitemOpen
  \bibfield  {author} {\bibinfo {author} {\bibfnamefont {J.}~\bibnamefont
  {Lee}}, \bibinfo {author} {\bibfnamefont {J.}~\bibnamefont {Zhang}}, \ and\
  \bibinfo {author} {\bibfnamefont {C.~C.}\ \bibnamefont {Lu}},\ }\bibfield
  {title} {\enquote {\bibinfo {title} {Incomplete {LU} preconditioning for
  large scale dense complex linear systems from electromagnetic wave scattering
  problems},}\ }\href@noop {} {\bibfield  {journal} {\bibinfo  {journal}
  {Journal of Computational Physics}\ }\textbf {\bibinfo {volume} {185}},\
  \bibinfo {pages} {158--175} (\bibinfo {year} {2003})}\BibitemShut {NoStop}%
\bibitem [{\citenamefont {Alves}, \citenamefont {Oliveira},\ and\ \citenamefont
  {Pinho}(2003)}]{alves2003convergent}%
  \BibitemOpen
  \bibfield  {author} {\bibinfo {author} {\bibfnamefont {M.~A.}\ \bibnamefont
  {Alves}}, \bibinfo {author} {\bibfnamefont {P.~J.}\ \bibnamefont {Oliveira}},
  \ and\ \bibinfo {author} {\bibfnamefont {F.~T.}\ \bibnamefont {Pinho}},\
  }\bibfield  {title} {\enquote {\bibinfo {title} {A convergent and universally
  bounded interpolation scheme for the treatment of advection},}\ }\href@noop
  {} {\bibfield  {journal} {\bibinfo  {journal} {International Journal for
  Numerical Methods in Fluids}\ }\textbf {\bibinfo {volume} {41}},\ \bibinfo
  {pages} {47--75} (\bibinfo {year} {2003})}\BibitemShut {NoStop}%
\bibitem [{\citenamefont {Khan}\ and\ \citenamefont
  {Sasmal}(2021)}]{khan2021elastic}%
  \BibitemOpen
  \bibfield  {author} {\bibinfo {author} {\bibfnamefont {M.~B.}\ \bibnamefont
  {Khan}}\ and\ \bibinfo {author} {\bibfnamefont {C.}~\bibnamefont {Sasmal}},\
  }\bibfield  {title} {\enquote {\bibinfo {title} {Elastic instabilities and
  bifurcations in flows of wormlike micellar solutions past single and two
  vertically aligned microcylinders: Effect of blockage and gap ratios},}\
  }\href@noop {} {\bibfield  {journal} {\bibinfo  {journal} {Physics of
  Fluids}\ }\textbf {\bibinfo {volume} {33}},\ \bibinfo {pages} {033109}
  (\bibinfo {year} {2021})}\BibitemShut {NoStop}%
\bibitem [{\citenamefont {Pakdel}\ and\ \citenamefont
  {McKinley}(1996)}]{pakdel1996elastic}%
  \BibitemOpen
  \bibfield  {author} {\bibinfo {author} {\bibfnamefont {P.}~\bibnamefont
  {Pakdel}}\ and\ \bibinfo {author} {\bibfnamefont {G.~H.}\ \bibnamefont
  {McKinley}},\ }\bibfield  {title} {\enquote {\bibinfo {title} {Elastic
  instability and curved streamlines},}\ }\href@noop {} {\bibfield  {journal}
  {\bibinfo  {journal} {Physical Review Letters}\ }\textbf {\bibinfo {volume}
  {77}},\ \bibinfo {pages} {2459} (\bibinfo {year} {1996})}\BibitemShut
  {NoStop}%
\bibitem [{\citenamefont {McKinley}, \citenamefont {Pakdel},\ and\
  \citenamefont {{\"O}ztekin}(1996)}]{mckinley1996rheological}%
  \BibitemOpen
  \bibfield  {author} {\bibinfo {author} {\bibfnamefont {G.~H.}\ \bibnamefont
  {McKinley}}, \bibinfo {author} {\bibfnamefont {P.}~\bibnamefont {Pakdel}}, \
  and\ \bibinfo {author} {\bibfnamefont {A.}~\bibnamefont {{\"O}ztekin}},\
  }\bibfield  {title} {\enquote {\bibinfo {title} {Rheological and geometric
  scaling of purely elastic flow instabilities},}\ }\href@noop {} {\bibfield
  {journal} {\bibinfo  {journal} {Journal of Non-Newtonian Fluid Mechanics}\
  }\textbf {\bibinfo {volume} {67}},\ \bibinfo {pages} {19--47} (\bibinfo
  {year} {1996})}\BibitemShut {NoStop}%
\bibitem [{\citenamefont {Pakdel}\ and\ \citenamefont
  {McKinley}(1998)}]{pakdel1998cavity}%
  \BibitemOpen
  \bibfield  {author} {\bibinfo {author} {\bibfnamefont {P.}~\bibnamefont
  {Pakdel}}\ and\ \bibinfo {author} {\bibfnamefont {G.~H.}\ \bibnamefont
  {McKinley}},\ }\bibfield  {title} {\enquote {\bibinfo {title} {Cavity flows
  of elastic liquids: purely elastic instabilities},}\ }\href@noop {}
  {\bibfield  {journal} {\bibinfo  {journal} {Physics of Fluids}\ }\textbf
  {\bibinfo {volume} {10}},\ \bibinfo {pages} {1058--1070} (\bibinfo {year}
  {1998})}\BibitemShut {NoStop}%
\bibitem [{\citenamefont {Haward}\ \emph {et~al.}(2019)\citenamefont {Haward},
  \citenamefont {Kitajima}, \citenamefont {Toda-Peters}, \citenamefont
  {Takahashi},\ and\ \citenamefont {Shen}}]{haward2019flow}%
  \BibitemOpen
  \bibfield  {author} {\bibinfo {author} {\bibfnamefont {S.~J.}\ \bibnamefont
  {Haward}}, \bibinfo {author} {\bibfnamefont {N.}~\bibnamefont {Kitajima}},
  \bibinfo {author} {\bibfnamefont {K.}~\bibnamefont {Toda-Peters}}, \bibinfo
  {author} {\bibfnamefont {T.}~\bibnamefont {Takahashi}}, \ and\ \bibinfo
  {author} {\bibfnamefont {A.~Q.}\ \bibnamefont {Shen}},\ }\bibfield  {title}
  {\enquote {\bibinfo {title} {Flow of wormlike micellar solutions around
  microfluidic cylinders with high aspect ratio and low blockage ratio},}\
  }\href@noop {} {\bibfield  {journal} {\bibinfo  {journal} {Soft Matter}\
  }\textbf {\bibinfo {volume} {15}},\ \bibinfo {pages} {1927--1941} (\bibinfo
  {year} {2019})}\BibitemShut {NoStop}%
\bibitem [{\citenamefont {Casanellas}\ \emph {et~al.}(2016)\citenamefont
  {Casanellas}, \citenamefont {Alves}, \citenamefont {Poole}, \citenamefont
  {Lerouge},\ and\ \citenamefont {Lindner}}]{casanellas2016stabilizing}%
  \BibitemOpen
  \bibfield  {author} {\bibinfo {author} {\bibfnamefont {L.}~\bibnamefont
  {Casanellas}}, \bibinfo {author} {\bibfnamefont {M.~A.}\ \bibnamefont
  {Alves}}, \bibinfo {author} {\bibfnamefont {R.~J.}\ \bibnamefont {Poole}},
  \bibinfo {author} {\bibfnamefont {S.}~\bibnamefont {Lerouge}}, \ and\
  \bibinfo {author} {\bibfnamefont {A.}~\bibnamefont {Lindner}},\ }\bibfield
  {title} {\enquote {\bibinfo {title} {The stabilizing effect of shear thinning
  on the onset of purely elastic instabilities in serpentine microflows},}\
  }\href@noop {} {\bibfield  {journal} {\bibinfo  {journal} {Soft matter}\
  }\textbf {\bibinfo {volume} {12}},\ \bibinfo {pages} {6167--6175} (\bibinfo
  {year} {2016})}\BibitemShut {NoStop}%
\bibitem [{\citenamefont {Kolmogorov}(1941)}]{kolmogorov1941degeneration}%
  \BibitemOpen
  \bibfield  {author} {\bibinfo {author} {\bibfnamefont {A.~N.}\ \bibnamefont
  {Kolmogorov}},\ }\bibfield  {title} {\enquote {\bibinfo {title} {On
  degeneration (decay) of isotropic turbulence in an incompressible viscous
  liquid},}\ }in\ \href@noop {} {\emph {\bibinfo {booktitle} {Doklady Akademii
  Nauk SSSR}}},\ Vol.~\bibinfo {volume} {31}\ (\bibinfo {year} {1941})\ pp.\
  \bibinfo {pages} {538--540}\BibitemShut {NoStop}%
\bibitem [{\citenamefont {Groisman}\ and\ \citenamefont
  {Steinberg}(2000)}]{groisman2000elastic}%
  \BibitemOpen
  \bibfield  {author} {\bibinfo {author} {\bibfnamefont {A.}~\bibnamefont
  {Groisman}}\ and\ \bibinfo {author} {\bibfnamefont {V.}~\bibnamefont
  {Steinberg}},\ }\bibfield  {title} {\enquote {\bibinfo {title} {Elastic
  turbulence in a polymer solution flow},}\ }\href@noop {} {\bibfield
  {journal} {\bibinfo  {journal} {Nature}\ }\textbf {\bibinfo {volume} {405}},\
  \bibinfo {pages} {53--55} (\bibinfo {year} {2000})}\BibitemShut {NoStop}%
\bibitem [{\citenamefont {Steinberg}(2021)}]{steinberg2021elastic}%
  \BibitemOpen
  \bibfield  {author} {\bibinfo {author} {\bibfnamefont {V.}~\bibnamefont
  {Steinberg}},\ }\bibfield  {title} {\enquote {\bibinfo {title} {Elastic
  turbulence: an experimental view on inertialess random flow},}\ }\href@noop
  {} {\bibfield  {journal} {\bibinfo  {journal} {Annual Review of Fluid
  Mechanics}\ }\textbf {\bibinfo {volume} {53}},\ \bibinfo {pages} {27--58}
  (\bibinfo {year} {2021})}\BibitemShut {NoStop}%
\bibitem [{\citenamefont {Varshney}\ and\ \citenamefont
  {Steinberg}(2017)}]{varshney2017elastic}%
  \BibitemOpen
  \bibfield  {author} {\bibinfo {author} {\bibfnamefont {A.}~\bibnamefont
  {Varshney}}\ and\ \bibinfo {author} {\bibfnamefont {V.}~\bibnamefont
  {Steinberg}},\ }\bibfield  {title} {\enquote {\bibinfo {title} {Elastic wake
  instabilities in a creeping flow between two obstacles},}\ }\href@noop {}
  {\bibfield  {journal} {\bibinfo  {journal} {Physical Review Fluids}\ }\textbf
  {\bibinfo {volume} {2}},\ \bibinfo {pages} {051301} (\bibinfo {year}
  {2017})}\BibitemShut {NoStop}%
\bibitem [{\citenamefont {Grilli}, \citenamefont {V{\'a}zquez-Quesada},\ and\
  \citenamefont {Ellero}(2013)}]{grilli2013transition}%
  \BibitemOpen
  \bibfield  {author} {\bibinfo {author} {\bibfnamefont {M.}~\bibnamefont
  {Grilli}}, \bibinfo {author} {\bibfnamefont {A.}~\bibnamefont
  {V{\'a}zquez-Quesada}}, \ and\ \bibinfo {author} {\bibfnamefont
  {M.}~\bibnamefont {Ellero}},\ }\bibfield  {title} {\enquote {\bibinfo {title}
  {Transition to turbulence and mixing in a viscoelastic fluid flowing inside a
  channel with a periodic array of cylindrical obstacles},}\ }\href@noop {}
  {\bibfield  {journal} {\bibinfo  {journal} {Physical Review Letters}\
  }\textbf {\bibinfo {volume} {110}},\ \bibinfo {pages} {174501} (\bibinfo
  {year} {2013})}\BibitemShut {NoStop}%
\bibitem [{\citenamefont {Yang}, \citenamefont {Yao},\ and\ \citenamefont
  {Wen}(2020)}]{yang2020experimental}%
  \BibitemOpen
  \bibfield  {author} {\bibinfo {author} {\bibfnamefont {H.}~\bibnamefont
  {Yang}}, \bibinfo {author} {\bibfnamefont {G.}~\bibnamefont {Yao}}, \ and\
  \bibinfo {author} {\bibfnamefont {D.}~\bibnamefont {Wen}},\ }\bibfield
  {title} {\enquote {\bibinfo {title} {Experimental investigation on convective
  heat transfer of shear-thinning fluids by elastic turbulence in a serpentine
  channel},}\ }\href@noop {} {\bibfield  {journal} {\bibinfo  {journal}
  {Experimental Thermal and Fluid Science}\ }\textbf {\bibinfo {volume}
  {112}},\ \bibinfo {pages} {109997} (\bibinfo {year} {2020})}\BibitemShut
  {NoStop}%
\bibitem [{\citenamefont {Li}\ \emph {et~al.}(2017)\citenamefont {Li},
  \citenamefont {Li}, \citenamefont {Zhang}, \citenamefont {Li}, \citenamefont
  {Qian},\ and\ \citenamefont {Joo}}]{li2017efficient}%
  \BibitemOpen
  \bibfield  {author} {\bibinfo {author} {\bibfnamefont {D.-Y.}\ \bibnamefont
  {Li}}, \bibinfo {author} {\bibfnamefont {X.-B.}\ \bibnamefont {Li}}, \bibinfo
  {author} {\bibfnamefont {H.-N.}\ \bibnamefont {Zhang}}, \bibinfo {author}
  {\bibfnamefont {F.-C.}\ \bibnamefont {Li}}, \bibinfo {author} {\bibfnamefont
  {S.}~\bibnamefont {Qian}}, \ and\ \bibinfo {author} {\bibfnamefont {S.~W.}\
  \bibnamefont {Joo}},\ }\bibfield  {title} {\enquote {\bibinfo {title}
  {Efficient heat transfer enhancement by elastic turbulence with polymer
  solution in a curved microchannel},}\ }\href@noop {} {\bibfield  {journal}
  {\bibinfo  {journal} {Microfluidics and Nanofluidics}\ }\textbf {\bibinfo
  {volume} {21}},\ \bibinfo {pages} {10} (\bibinfo {year} {2017})}\BibitemShut
  {NoStop}%
\end{thebibliography}%

\end{document}